\documentclass [11pt]{article}
 \pdfoutput=1
\usepackage{amsfonts}
\usepackage{graphicx}
\usepackage{amsmath,amssymb}
\usepackage{latexsym}
\usepackage{mathrsfs}
\usepackage{bbold}
\usepackage{color}
\usepackage{slashed}
\definecolor{red}{rgb}{1,0,0}
\usepackage{cancel}
\usepackage{comment}
\usepackage[nosort]{cite}
\usepackage{color}
\usepackage{multirow}

\usepackage{soul} 

\usepackage{setspace} 

\definecolor{red}{rgb}{1,0,0}

\makeatletter
\def\section{\@startsection {section}{1}{\z@}{-3.5ex plus -1ex minus
 -.2ex}{2.3ex plus .2ex}{\large\bf}}
\def\subsection{\@startsection{subsection}{2}{\z@}{-3.25ex plus -1ex
minus -.2ex}{1.5ex plus .2ex}{\normalsize\bf}}
\makeatother
\makeatletter

\@addtoreset{equation}{section}

\makeatother

\def\bel{\begin{equation}\begin{aligned}}
\def\eel{\end{aligned}\end{equation}}

\def\bea{\begin{eqnarray}} \def\eea{\end{eqnarray}}
\def\be{\begin{equation}} \def\ee{\end{equation}} 
\def\nn{\nonumber}

\newcommand{\promille}{%
  \relax\ifmmode\promillezeichen
        \else\leavevmode\(\mathsurround=0pt\promillezeichen\)\fi}
\newcommand{\promillezeichen}{%
  \kern-.05em%
  \raise.5ex\hbox{\the\scriptfont0 0}%
  \kern-.15em/\kern-.15em%
  \lower.25ex\hbox{\the\scriptfont0 00}}

\usepackage[colorlinks,linkcolor=black,citecolor=blue,urlcolor=blue,linktocpage]{hyperref}

\begin{document}

\thispagestyle{empty}

\begin{center}

\vspace*{-.6cm}

\hfill SISSA 02/2016/FISI \\

\begin{center}

\vspace*{1.1cm}

{\Large\bf\centerline{Seed Conformal Blocks  in 4D CFT}} 

\end{center}

\vspace{0.8cm}
{\bf Alejandro Castedo Echeverri$^{a}$, Emtinan Elkhidir$^{a}$,\\[3mm]}
{\bf  Denis Karateev$^{a}$, Marco Serone$^{a,b}$}

\vspace{1.cm}

${}^a\!\!$
{\em SISSA and INFN, Via Bonomea 265, I-34136 Trieste, Italy}

\vspace{.1cm}

${}^b\!\!$
{\em ICTP, Strada Costiera 11, I-34151 Trieste, Italy}

\end{center}

\vspace{1cm}

\centerline{\bf Abstract}
\vspace{2 mm}
\begin{quote}

We compute in closed analytical form the minimal set of ``seed" conformal blocks associated to the exchange of
generic mixed symmetry spinor/tensor operators in an arbitrary representation $(\ell,\bar \ell)$ of the Lorentz group in four dimensional conformal field theories.
These blocks arise from 4-point functions involving two scalars, one $(0,|\ell-\bar \ell|)$ and one $(|\ell-\bar \ell|,0)$ spinors or tensors. 
We directly solve the set of Casimir equations, that can elegantly be written in a compact form for any $(\ell,\bar \ell)$,
by using an educated ansatz and reducing the problem to an algebraic linear system. 
Various details on the form of the ansatz  have been deduced by using the so called shadow formalism.
The complexity of the conformal blocks depends on the value of $p=|\ell-\bar \ell |$ and grows with $p$, in analogy to what happens
to scalar conformal blocks in $d$ even space-time dimensions as $d$ increases.
These results open the way to bootstrap  4-point functions involving arbitrary spinor/tensor operators in four dimensional conformal field theories.

\end{quote}

\newpage

\tableofcontents

\section{Introduction}

There has recently been a revival of interest in the old idea of the bootstrap program \cite{Ferrara:1973yt,Polyakov:1974gs} after ref.\cite{Rattazzi:2008pe} observed that its applicability extends
to Conformal Field Theories (CFTs) in more than two space-time dimensions (2D). Since ref.\cite{Rattazzi:2008pe}, several constraints have been imposed on CFTs data, namely  spectrum of operators and 
Operator Product Expansion (OPE) coefficients, in CFTs in different dimensions, up to 6D (see e.g. ref. \cite{Beem:2015aoa}).
Imposing additional (mild and reasonable) assumptions, one can also compute CFT data of given CFTs, the 3D Ising model being
probably the most striking example \cite{ElShowk:2012ht,El-Showk:2014dwa,Simmons-Duffin:2015qma}.

The bootstrap approach is a systematic way of imposing crossing symmetry in correlation functions. Analyzing 4-point functions is enough to get constraints on the CFT data. In order to be able to implement the bootstrap program, it is essential to be able to decompose the 4-point functions in terms
of the individual contributions of the infinite number of primary operators (and all their descendants) that can be exchanged in any given kinematical channel, namely that appear in the OPE
of the four external operators.
For each primary operator, in particular, one has to resum the infinite series of associated descendant operators in what is called a Conformal Partial Wave (CPW).
CPWs can be decomposed in terms of scalar functions known as Conformal Blocks (CBs).
Before the advent of ref.\cite{Rattazzi:2008pe}, the only known CBs were those associated to symmetric traceless tensors exchanged in scalar 4-point functions in even number of dimensions \cite{Dolan:2000ut,Dolan:2003hv},
denoted for short scalar symmetric CBs in the following.

Not surprisingly, after ref.\cite{Rattazzi:2008pe} significant progress has been made in computing CBs.
Various techniques have been introduced to determine in an expanded form the scalar symmetric CBs in 3D \cite{ElShowk:2012ht,Hogervorst:2013sma,Hogervorst:2013kva,Kos:2013tga}, where a general closed analytic expression has not been found so far. In particular, using the techniques of ref.\cite{Kos:2013tga} and the further developments in ref.\cite{Penedones:2015aga}, CBs associated to the exchange of fermion operators in 3D have recently been computed \cite{Iliesiu:2015akf}.
In ref.\cite{Costa:2011dw} it has been shown how to relate, in any number of dimensions, symmetric CBs in correlators of external traceless symmetric operators to the known scalar symmetric blocks. 
In ref.\cite{SimmonsDuffin:2012uy} the so called shadow formalism method  \cite{Ferrara:1972xe,Ferrara:1972uq,Ferrara:1972ay,Ferrara:1973vz}, already used in ref.\cite{Dolan:2000ut}, has been further developed to compute any CB in any number of dimensions. Although very powerful, the shadow formalism leads to quite involved and not so  enlightening expressions. Applications of this method for some specific correlators appeared in ref.\cite{Rejon-Barrera:2015bpa}. 
Some other limits of the known CBs have been discussed in refs.\cite{Fitzpatrick:2013sya,Behan:2014dxa}, as well as  their interpretation in terms of Witten diagrams in Anti de Sitter (AdS) space  \cite{Hijano:2015zsa}.
Despite significant progress, not much has been done in the analysis of CBs associated to mixed symmetry tensor (or fermion) operators,  denoted simply mixed tensor CBs in the following.
Such CBs are crucial to extend the bootstrap program to tensor correlators in CFTs in $d>3$ space-time dimensions, where such operators can appear in the OPE between two external fields.\footnote{In 3D, scalar blocks, the recently computed fermion ones \cite{Iliesiu:2015akf} and the results of ref.\cite{Costa:2011dw} allow us to determine any other CB.}

Mixed tensor CBs in 4D CFTs have recently been analyzed in ref.\cite{Echeverri:2015rwa}. In particular it has been shown there how to relate, by means of differential operators,  mixed tensor CBs appearing in an arbitrary
spinor/tensor 4-point correlator (not necessarily traceless symmetric operators) to a basis of minimal mixed tensor CBs. These ``seed" blocks arise from 4-point functions involving two scalars
and two tensor fields in the $(0,p)$ and $(p,0)$ representations of the Lorentz group, with $p$ an arbitrary integer ($(1,0)$ is a fermion). 
Such 4-point functions are the simplest ones (i.e. with the least number of tensor structures) where $(\ell+p,\ell)$ or $(\ell,\ell+p)$ mixed symmetry (bosonic or fermionic) tensors can be exchanged in some OPE limit, for any $\ell$. 

The aim of this paper is to compute the ``seed" CBs identified in ref.\cite{Echeverri:2015rwa}. 
We will be able to find in closed analytical form the set of seed CBs associated to the exchange of operators in the $(\ell,\bar \ell)$ representations of the Lorentz group.
They are labelled by the positive integer $p=|\ell-\bar \ell|$ and are thus infinite. We consider at the same time CBs associated to both bosonic (even $p$) and fermionic (odd $p$) mixed symmetry tensor operators.
For each given $p$, one has to determine $p+1$ CBs $G^{(p)}_e$, $e=0,1,\ldots,p$, one for each tensor structure appearing in 
the corresponding CPW.  Using the 6D embedding formalism in twistor space in index-free notation \cite{SimmonsDuffin:2012uy,Elkhidir:2014woa}, we will 
be able to write in a compact form the system of Casimir equations satisfied by the $p+1$ CBs.
Solving the Casimir system is a hard task, that also requires the knowledge of some boundary conditions, like the asymptotic behaviour of $G^{(p)}_e$.
We first attack the problem using the shadow formalism. With the use of some tricks, we find integral expressions of $G^{(p)}_e$ for any $p$ and $\ell$, and explicit
expressions for $p=1,2$ (and any $\ell$).
The shadow formalism also allows us to get the asymptotic behaviour of $G^{(p)}_e$ in the OPE limit $u\rightarrow 0$, $v\rightarrow 1$ for $p=1,2$ and any $\ell$, and for any $p$ and $\ell=0$, together with
some other information on the structure of the blocks.
Thanks to the knowledge acquired in this way, we will be able to go back to the Casimir system and solve it for any $p$ and $\ell$,
using generalizations of the methods introduced  in ref.\cite{Dolan:2003hv} (and further refined in ref.\cite{Dolan:2011dv}) to compute 6D symmetric CBs for scalar correlators.
Like scalar blocks in higher even dimensions, the mixed tensor CBs are found using an ansatz given by a sum of hyper-geometric functions with unknown coefficients $c_{m,n}^e$.
In this way a system of $p+1$ linear coupled differential equations of second order in two variables is reduced to an algebraic linear system for $c_{m,n}^e$.
The set of non-trivial coefficients $c_{m,n}^e$, determined by solving the linear system, admits a useful geometric interpretation. They span a two-dimensional lattice in the $(m,n)$ plane. 
For each CB labelled by $e$, the shape of the lattice is an octagon, with $p$ and $e$ dependent edges. For large $p$, 
the total number of coefficients $c_{m,n}^e$ grows like $p^3$ and their explicit form becomes more and more complicated as $p$ increases.
We point out that a similar geometric interpretation applies also to the set of non-trivial coefficients $x_{m,n}$ entering the solution for the symmetric scalar blocks in even number of dimensions.

The structure of the paper is as follows. In section 2 we briefly review the results of ref.\cite{Echeverri:2015rwa} and define the CPWs and the CBs of interest.
In section 3, we derive the system of $p+1$ Casimir equations satisfied by the CBs $G^{(p)}_e$, for any $p$. This is reported in eq.(\ref{eq:CasimirSystem}).
In section 4 we compute the CPWs using the shadow formalism approach. In particular, in subsection 4.1 we derive compact integral expressions of the CBs for any $p$ and $\ell$, eq.(\ref{eq:MasterRelation}).
In subsection 4.2 we write a more explicit expression of the CBs for $\ell=0$ and in subsection 4.3 we find another integral expression for the CPW, eq.(\ref{FinalShadow}), more suitable to perform
computations with $\ell\neq 0$. The solution of the Casimir system of equations is described in section 5. In subsection 5.1 we derive, by extending the results found in section 4, the asymptotic
behaviour of the CBs. We discuss the form of the ansatz in subsection 5.2, and finally we reduce the coupled differential equations to an algebraic system in subsection 5.3.
The solution of the CBs is finally derived in subsection 5.4, eq.(\ref{eq:Ansatz}). 
In subsection 5.5 we draw an analogy between the mixed tensor blocks $G_e^{(p)}$ and the symmetric scalar blocks in $d$ even dimensions.
We conclude in section 6. 
Various technical details, as well as the explicit form of the coefficient defining the fermionic CBs entering scalar-fermion correlators ($p=1$) are reported in two appendices.

The explicit form of all the coefficients $c_{m,n}^e$  entering the CBs (\ref{eq:Ansatz}) for $p=1,2,3,4$ can be downloaded from \href{https://sites.google.com/site/dskarateev/downloads}
{https://sites.google.com/site/dskarateev/downloads}.

\section{Deconstructing Conformal Partial Waves}
\label{sec:seed}

In 4D CFTs, for a given 4-point function, CBs and CPWs are labelled by the quantum numbers of the exchanged primary operator and thus they depend
on its scaling dimension $\Delta$ and representation $(\ell,\bar \ell)$ of the 4D Lorentz group, with $\ell$ and $\bar \ell$ positive integers.
Four-point functions involving only scalar fields are the best known. In any channel, the exchanged operators have $\bar \ell = \ell$, i.e.
they are all and only traceless symmetric tensors.
In this case CPW and CB are equivalent up to a kinematic factor and their analytic form has been derived in a remarkable compact form in refs.\cite{Dolan:2000ut,Dolan:2003hv} for any $\Delta$ and $\ell$.
Four-point functions involving tensor (or fermion) operators are considerably more complicated because different tensor structures arise and more operators can be exchanged.
A generic fermion-tensor 4-point function can be parametrized as 
\be
\langle {\cal O}^{I_1}_1(x_1) {\cal O}^{I_2}_2(x_2) {\cal O}^{I_3}_3(x_3) {\cal O}^{I_4}_4(x_4)\rangle = \mathcal{K}_4 \sum_{n=1}^{N_4} g_n(u,v) {\cal T}_n^{I_1I_2I_3I_4}(x_i) \,,
\label{Gen4pt}
\ee
where $I_{i}$ are schematic Lorentz indices of the operators ${\cal O}_i(x_i)$,  
\be\label{eq:4kinematic}
\mathcal{K}_4 = \bigg(\frac{x_{24}^2}{x_{14}^2}\bigg)^{\frac{\tau_1-\tau_2}2}
\bigg(\frac{x_{14}^2}{x_{13}^2}\bigg)^{\frac{\tau_3-\tau_4}{2}}
(x_{12}^2)^{-\frac{\tau_1+\tau_2}2} (x_{34}^2)^{-\frac{\tau_3+\tau_4}2} 
\ee
is a kinematic factor, $x_{ij}^2=(x_i-x_j)_\mu(x_i-x_j)^\mu$, $\tau_i=\Delta_i +(\ell_i+\bar \ell_i)/2$, 
$u$ and $v$ are the usual conformally invariant cross ratios
\be
u=\frac{x_{12}^2x_{34}^2}{x_{13}^2x_{24}^2}\,, \ \ \ v=\frac{x_{14}^2 x_{23}^2}{x_{13}^2x_{24}^2}\,,
\label{uv4d}
\ee
and ${\cal T}_n^{I_1I_2I_3I_4}(x_i)$ are kinematically determined tensor structures. 
The dynamical information of the 4-point function is encoded in the $N_4$ functions $g_n(u,v)$. As we mentioned, a bootstrap analysis requires 
to rewrite  the 4-point function (\ref{Gen4pt}) in terms of the operators exchanged in any channel.
In the s-channel (12-34), for instance,  we have
\be
\langle {\cal O}^{I_1}_1(x_1) {\cal O}^{I_2}_2(x_2) {\cal O}^{I_3}_3(x_3) {\cal O}^{I_4}_4(x_4)\rangle =  \sum_{i,j}\sum_{{\cal O}_r} \lambda_{{\cal O}_1{\cal O}_2{\cal O}_r}^i \lambda_{\bar {\cal O}_{\bar r}{\cal O}_3{\cal O}_4}^j W_{{\cal O}_1{\cal O}_2{\cal O}_3{\cal O}_4,{\cal O}_r}^{(i,j)I_1I_2I_3I_4}(x_i)\,,
\label{sch4pt}
\ee
where $i$ and $j$ run over the possible independent tensor structures associated to the three point functions $\langle {\cal O}_1{\cal O}_2{\cal O}_r\rangle$ and $\langle {\bar {\cal O}}_{\bar r}{\cal O}_3{\cal O}_4\rangle$, $\lambda$'s being their corresponding structure constants and $W_{{\cal O}_1{\cal O}_2{\cal O}_3{\cal O}_4}^{(p,q)I_1I_2I_3I_4}(u,v)$ are the associated CPWs. 
The sum over the exchanged primary operators ${\cal O}_r$ includes a sum over all possible representations $(\ell,\bar \ell)$ that can appear in the 4-point function and, for each representation, a sum over
all the possible primaries, i.e. a sum over all  possible scaling dimensions $\Delta_{{\cal O}_r}$. It is useful to define $\delta = |\bar \ell - \ell|$
and rearrange the sum over $(\ell,\bar \ell)$ in a sum over, say, $\ell$ and $\delta$. There is an important difference between these two sums. For any 4-point function, the sum over $l$ extends up to infinity, while the sum over $\delta$ is always finite.
More precisely, we have
\be\begin{split}
\delta= & \; 0,\,2 \,,\, \ldots \,,\,p-2,\, p, \ \ \   \ \ {\cal O}_r  \ \  {\rm bosonic}  \\
\delta= & \; 1,\,3 \,,\, \ldots \,,\,p-2,\, p,\ \ \ \  \ {\cal O}_r \ \ {\rm fermionic}.
\end{split}
\label{deltaDef}
\ee
In both cases, the integer $p$ is defined to be
\be
p = {\rm min} (\ell_1+\bar \ell_1 +  \ell_2+\bar \ell_2  , \ell_3+\bar \ell_3 +  \ell_4+\bar \ell_4)\,,
\label{pDef}
\ee
and is automatically an even or odd integer  when  ${\cal O}_r$ is a boson or a fermion operator.
There are several CPWs for each exchanged primary operator ${\cal O}_r$, depending on the number of allowed 3-point function structures.
They admit a parametrization like the 4-point function itself,
\be
W_{{\cal O}_1{\cal O}_2{\cal O}_3{\cal O}_4,{\cal O}_r}^{(i,j)I_1I_2I_3I_4}(x_i) = \mathcal{K}_4 \sum_{n=1}^{N_4} {g}_{{\cal O}_r,n}^{(i,j)}(u,v)  {\cal T}_n^{I_1I_2I_3I_4}(x_i) \,,
\label{WGen}
\ee
where ${g}^{(i,j)}_{{\cal O}_r,n}(u,v)$ are the CBs, scalar functions of $u$ and $v$ that depend on the dimensions and spins of the external and exchanged operators.
Imposing crossing symmetry by requiring  the equality of different channels is the essence of the bootstrap approach.
In order to successfully bootstrap the correlator (\ref{Gen4pt}), it is necessary to know the explicit form of the CPWs (\ref{WGen}), in particular the CBs  ${g}^{(i,j)}_{{\cal O}_r,n}(u,v)$.

It has been shown in ref. \cite{Echeverri:2015rwa} that the CPWs associated to an operator $\mathcal O^{(\ell,\ell+p)}$  (and similarly for its conjugate $\overline{\mathcal O}^{(\ell+p,\ell)}$) exchanged in the OPE channel $(12)(34)$  of a 4-point function 
$\langle \mathcal O_1 \mathcal O_2 \mathcal O_3 \mathcal O_4 \rangle$, can be obtained from a single CPW $W^{see
 d}_{\mathcal O^{(\ell,\ell+p)}}$ as follows:
\be\label{deconstructing}
W^{(i,j)}_{\mathcal O_1 \mathcal O_2 \mathcal O_3 \mathcal O_4,\mathcal O^{(\ell,\ell+p)}}= \mathcal D^i_{12}\mathcal D^j_{34} W^{seed}_{\mathcal O^{(\ell,\ell+p)}}\,,
\ee
where $\mathcal D^i_{12}$ and $\mathcal D^i_{34}$ are differential operators that depend on ${\cal O}_{1,2}$ and ${\cal O}_{3,4}$, respectively. For
even integer $p=2n$, the seed CPWs are those associated to 4-point functions of two scalar fields with one $(2n,0)$ and one $(0,2n)$ bosonic operators, 
while for odd integer $p=2n+1$,  they consist of 4-point functions of two scalar fields with one $(2n+1,0)$ and one $(0,2n+1)$ fermionic operators:\footnote{Strictly speaking, we focused in ref.\cite{Echeverri:2015rwa}
on the even $p$ case, but it is obvious that the same result applies to odd $p$.}
\bea\label{4dboson}
&& \langle \phi_1 (x_1) F_{2,\alpha_1 \alpha_2\dots \alpha_{2n}}(x_2) \phi_3 (x_3) \overline F^{\dot{\beta}_1\dot{\beta}_2 \dots \dot{\beta}_{2n}}_4(x_4) \rangle
 \,, \ \ \ \ \ \ \ \; p=2n \,, \\
\label{4dfermion} 
&&\langle \phi_1 (x_1) \psi_{2,\alpha_1 \alpha_2\dots \alpha_{2n+1}}(x_2) \phi_3 (x_3) \overline \psi^{\dot{\beta}_1\dot{\beta}_2 \dots \dot{\beta}_{2n+1}}_4(x_4) \rangle \,, \ \ \ p=2n+1\,.
\eea
In the above correlators, in the OPE channel $\langle(12)(34) \rangle$ primary operators ${\cal O}^{(\ell,\ell+\delta)}$ and their conjugates $\overline{\cal O}^{(\ell+\delta,\ell)}$ can be exchanged only with the values of $\delta$ indicated in eq.~(\ref{deltaDef}) and any $\ell$. 
There are several 4-point functions in which the operators  ${\cal O}^{(\ell,\ell+p)}$ and $\overline{\cal O}^{(\ell+p,\ell)}$ are exchanged and in which the corresponding CPWs have a unique structure.
Among these, the correlators (\ref{4dboson}) and (\ref{4dfermion}) are the ones with the minimum number of tensor structures and hence the simplest.
This is understood by noticing that for {\it any} value of $\delta$ (and not only for $\delta=p$) the operators ${\cal O}^{(\ell,\ell+\delta)}$ and their conjugates $\overline{\cal O}^{(\ell+\delta,\ell)}$ appear in both the $(12)$ and $(34)$ OPE's with one tensor structure only, since there is only one tensor structure in the corresponding three-point functions:
\bea
&&\langle \phi(x_1) F_{\alpha_1 \dots \alpha_{2n}}(x_2) \mathcal O^{\dot{\beta}_1 \dots \dot{\beta}_{\ell+\delta}}_{\alpha_1\dots \alpha_\ell}(x_0) \rangle\,, \ \ \ \ \; \langle \overline{\mathcal O}^{\dot{\beta}_1 \dots \dot{\beta}_\ell}_{\alpha_1\dots \alpha_{\ell+\delta}}(x_0)\phi(x_3) \overline F^{\dot{\beta}_1 \dots \dot{\beta}_{2n}}(x_4) \rangle \,,  \\
&&\langle \phi(x_1) \psi_{\alpha_1 \dots \alpha_{2n+1}}(x_2) \mathcal O^{\dot{\beta}_1 \dots \dot{\beta}_{\ell+\delta}}_{\alpha_1\dots \alpha_\ell}(x_0) \rangle\,, \ \ \langle \overline{\mathcal O}^{\dot{\beta}_1 \dots \dot{\beta}_\ell}_{\alpha_1\dots \alpha_{\ell+\delta}}(x_0)\phi(x_3) \overline \psi^{\dot{\beta}_1 \dots \dot{\beta}_{2n+1}}(x_4) \rangle \,.
\eea
This implies then that the number of 4-point tensor structures appearing in eqs.(\ref{4dboson}) and (\ref{4dfermion}) is the minimum possible and equals to $N_4=p+1$.

Summarizing,  the problem of computing CPWs and CBs associated to the exchange of mixed symmetry operators ${\cal O}^{(\ell,\ell+p)}$ and $\overline{\cal O}^{(\ell+p,\ell)}$ in any 4-point function is reduced to the computation of the $p+1$ CBs appearing in the decomposition of $W^{seed}_{\mathcal O^{(\ell,\ell+p)}}$ and $\overline W^{seed}_{\mathcal O^{(\ell+p,\ell)}}$.

Despite this simplification, the above computation is still technically challenging. A further great simplification occurs by using the 6D embedding formalism \cite{Dirac:1936fq,Mack:1969rr,Ferrara2,Dobrev:1977qv} 
in twistor space with index-free notation \cite{SimmonsDuffin:2012uy}. 
As we will see, among other things, this formalism spare us from explicitly writing tensor structures with open indices for the correlators (\ref{4dboson}) and (\ref{4dfermion}).
The 4D conformal group is isomorphic to the 6D Lorentz group $SO(4,2)$, so by embedding the 4D fields $\phi(x)$ into 6D counterparts $\Phi (X)$, the non-linear conformal transformations turn 
into linear 6D Lorentz transformations. 6D fields are defined on a 4 dimensional subspace: the projective ($X\sim \lambda X$) light-cone ($X_M X^M=0$) (see e.g. refs.\cite{Weinberg:2010fx,Costa:2011mg} for further details).
Using the local isomorphism $SO(4,2) \sim SU(2,2)$, 4D Weyl spinors $\psi_\alpha$(x)  can be embedded either into twistors $\Psi_a(X)$ subject to a transversality constraint \cite{Weinberg:2010fx} or to twistors $\overline\Psi^b(X)$ 
subject to a gauge redundancy \cite{SimmonsDuffin:2012uy}. Following refs.\cite{SimmonsDuffin:2012uy,Elkhidir:2014woa}, we adopt here the latter possibility.
A general 4D primary field $\mathcal O^{\dot{\beta}_1\dots\dot{\beta}_{\bar \ell}}_{\alpha_1\dots\alpha_\ell} (x)$, with scaling dimension $\Delta$ in the $(\ell,\bar \ell)$ representation is embedded in a 6D multi-twistor field $O_{b_1\dots b_{\bar \ell}}^{a_1\dots a_\ell}(X)$,  homogeneous in $X$ with degree $\tau=\Delta+(\ell+\bar \ell)/2$.
We can saturate the indices of multi-twistor fields by multiplying them by auxiliary twistors  $S$ and $\overline S $ to get index-free scalar quantities:
\be
O^{(\ell,\bar \ell)}(X,S,\overline S) = S_{a_1}\dots S_{a_\ell} \overline S^{b_1}\dots \overline S^{b_{\bar \ell}}O_{b_1\dots b_{\bar \ell}}^{a_1\dots a_\ell}(X) \,.
\ee 
The gauge redundancy requires that effectively
\begin{equation}\label{gaugered}
\overline{\textbf X}^{b a}S_a=\overline S^a\textbf X_{a b}=\overline S^a S_a=0\,,
\end{equation}
where $\textbf{X}_{ab}=X_{M}\Sigma^M_{ab}$, $\overline{\textbf{X}}^{ab}=X^{M}\overline\Sigma_M^{ab}$ , $\Sigma^M$ and $\overline\Sigma_M$ are the 6D chiral gamma matrices. The light-cone condition requires 
also $\overline{\textbf X}\textbf X=0$  (for all definitions and more details see ref.\cite{Elkhidir:2014woa}).
All tensor structures in twistor correlators can be written in terms of scalar functions of auxiliary twistors  $S$'s and $\overline S $'s. For $n$-point correlators one can find a basis of all possible linearly independent functions of $(S_1 ,\dots , S_n ,\overline S_1,\dots ,\overline S_n)$; for $n=4$ such basis includes, among others, the following invariants ($i\neq j \neq k \neq l$):
 \be
 I_{ij} \equiv \overline S_i S_j ,\ \ \ \ \ \
 J_{ij,kl} \equiv \frac{\overline S_i \textbf X_k \overline{\textbf X_l} S_j}{X_{kl}} ,
 \label{IijJijkl}
 \ee
 where $X_{ij}=X_i^M X_{jM}$. An independent basis for the $p+1$ tensor structures appearing in the 6D uplift of the correlators (\ref{4dboson}) and (\ref{4dfermion}) can be obtained from 
 the invariants in eq.(\ref{IijJijkl}):
 \be\label{our seed corr}
\langle \Phi_1(X_1)F_2^{(p,0)}(X_2,S_2)\Phi_3(X_3)\overline F_4^{(0,p)}(X_4,\overline S_4)\rangle =\mathcal K_4 \sum_{n=0}^{p}g_n (U,V) I_{42}^n J_{42,31}^{p-n}\,,
\ee
where $\mathcal K_4$, $U$ and $V$ are the 6D analogues of eqs.(\ref{eq:4kinematic})-(\ref{uv4d}), obtained by replacing $x_{ij}^2\rightarrow X_{ij}$.
We denote the 6D seed CPW associated to the exchange of the fields $O^{(\ell,\ell+p)}$ and $\overline{O}^{(\ell+p,\ell)}$ in the 4-point function (\ref{our seed corr}) by $W^{seed}(p)$ and $\overline W^{seed}(p)$, respectively. 
They are parametrized in terms of $p+1$ CBs as follows:
\be\label{eq:CPW_parametrization}\begin{aligned}
W^{seed}(p)\; =& \; \;\mathcal K_4 \sum_{e=0}^{p}G^{(p)}_e (U,V) I_{42}^e J_{42,31}^{p-e},\\
\overline W^{seed}(p) \;=&\;\;\mathcal K_4 \sum_{e=0}^{p}\overline G^{(p)}_e (U,V) I_{42}^e J_{42,31}^{p-e}.
\end{aligned}\ee
For simplicity, we have dropped in eq.(\ref{eq:CPW_parametrization}) the dependence of $G^{(p)}_e$ and $\overline G^{(p)}_e$ on $\Delta$ and $\ell$. The CBs depend also on the external operator dimensions, more precisely on $a$ and $b$, defined as 
\be\label{abDef}
a\equiv\frac{\tau_2-\tau_1}{2}=\frac{\Delta_2-\Delta_1}{2}+\frac{p}{4}\,, \ \  \ \  \ \  \ \  b\equiv\frac{\tau_3-\tau_4}{2}=\frac{\Delta_3-\Delta_4}{2}-\frac{p}{4}\,.
\ee 
For simplicity of notation, we no longer distinguish between even and odd values of $p$, since we can consider both cases simultaneously. It is then 
understood that in the corrrelator (\ref{our seed corr}) $F_2^{(p,0)}$ and $\overline F_4^{(0,p)}$ are 6D uplifts of 4D fermion fields for $p$ odd.

It is possible to get $W^{seed}(p)$ from $\overline W^{seed}(p)$, or vice versa, using the results of ref.\cite{Echeverri:2015rwa} and a parity transformation $\mathcal P$.
We have
\be \label{WbarDiff}
\overline W^{seed}(p)= \mathcal P \,\, W_{\Phi_1\overline F_2\Phi_3F_4,O^{(\ell,\ell+p)}} \,,
\ee
where 
\be\label{WbarFromW}
W_{\Phi_1\overline F_2\Phi_3F_4,O^{(\ell,\ell+p)}}=\frac{1}{2^{2p} \,( p!)^2}\Big( \prod_{n=1}^p c_n\Big) \ (\nabla_{12}\bar d_1 \widetilde D_1)^p(\nabla_{43}d_3 \widetilde D_3)^p W^{seed}(p)\Big|_{a\rightarrow a-\frac{p}{2} ,\, b\rightarrow b+\frac{p}{2}}
\ee
is the CPW associated to the parity dual 4-point function $\langle \Phi_{1}\overline F_2^{(0,p)}\Phi_3F_4^{(p,0)}\rangle$, and
\be
(c_n)^{-1}=(4+3p-2a-\tau-2n)(4+3p+2b-\tau-2n)\,, \ \ \tau = \Delta+\ell+\frac p2\,.
\ee
We do not report here the explicit form of the differential operators $\nabla_{ij} , \widetilde D_i , d_3 $ and $\bar d_1$, as well as the action of parity on them and on the 6D $SU(2,2)$ invariants, that
can be found  in ref. \cite{Echeverri:2015rwa}.
In fact, we will not use eq.(\ref{WbarDiff}) to compute $\overline W^{seed}(p)$, because
we will find an easier way to directly compute both $W^{seed}(p)$ and $\overline W^{seed}(p)$.

Instead of eq.(\ref{our seed corr}), we could have considered the alternative 4-point function
\be\label{34 permutation}
\langle \Phi_1(X_1)F_2^{(p,0)}(X_2)\overline F_3^{(0,p)}(X_3)\Phi_4(X_4)\rangle 
\ee
to calculate an analogue seed CPW $\widetilde W^{seed}(p)$. Since eq.(\ref{34 permutation}) is equal to eq.(\ref{our seed corr}) under the permutation $3\leftrightarrow4$, 
the CBs appearing in the decomposition of $W^{seed}{(p)}$ and $\widetilde W^{seed}(p)$ are related as follows:
\be
\widetilde G^{(p)}_e(U,V;a,b) =V^{a} G^{(p)}_e\Big(\frac UV,\frac 1V;a,-b\Big)\,, \ \ \ e=0,\ldots, p \,.
\ee 
The 4D CPWs $W^{seed}_{\mathcal O^{(\ell,\ell+p)}}$ and $\overline W^{seed}_{\mathcal O^{(\ell+p,\ell)}}$ are obtained by projecting to 4D their 6D counterparts $W^{seed}(p)$ and $\overline W^{seed}(p)$.
There is no need to explicitly perform such projection, because the 4D CBs are directly identified with their 6D counterparts. One has simply
\be
G^{(p)}_e(U,V)  = G^{(p)}_e(u,v) \,, \ \ \ \overline{G}^{(p)}_e(U,V)  = \overline{G}^{(p)}_e(u,v) \,,
\ee
where $G^{(p)}_e(u,v)$ and $ \overline{G}^{(p)}_e(u,v)$ are the 4D CBs entering  the r.h.s. of eq.(\ref{WGen}) when expanding the 4D CPWs $W^{seed}_{\mathcal O^{(\ell,\ell+p)}}$ and $\overline{W}^{seed}_{\mathcal O^{(\ell+p,\ell)}}$.

\section{The System of  Casimir Equations}
\label{sec:Casimir}

In this section we derive the system of second order Casimir equations for the seed conformal blocks defined in eq.~(\ref{eq:CPW_parametrization}).
Before addressing the more complicated case of interest, let us recall how the Casimir equation for scalar correlators is derived.
One starts by considering the 4-point function
\be \label{Ccomm}
\langle [\hat C,\phi_1(x_1)\phi_2(x_2)] \phi_3(x_3) \phi_4(x_4) \rangle\,,
\ee 
where $\hat C$ is the quadratic Casimir operator.\footnote{CBs satisfy also higher order equations obtained by means of higher Casimir invariants. We will not consider them in this paper, since
the quadratic Casimir will be enough for us to find the CB's. Here and in what follows we use a hat to denote an operator in the Hilbert space and to distinguish it from its explicit form in terms of differential operators, where no hat appears.}  
Recasting the generators of the 4D conformal group in a 6D form as $\hat L_{MN}$, with $M,N$ 6D indices, we have
\be\label{eq:casimir_operator}
\hat C =  \frac 12 \hat L_{MN} \hat L^{MN} \,.
\ee
The Casimir equation is derived by expressing eq.(\ref{Ccomm}) in two different ways. 
On one hand, we can replace in eq.(\ref{eq:casimir_operator}) the operator $\hat L_{MN}$ with its explicit action in terms of differential operators acting on the scalar fields inserted at the points $x_1$ and $x_2$:
$[\hat L_{MN},\phi(x)] = L_{MN}(x,\partial) \phi(x)$.
On the other hand, we might consider the (12) OPE. Scalar operators can only exchange symmetric traceless operators, so $p=0$ in this case, and one has  
\be\label{phiOPE}
\phi_1(x_1) \phi(x_2)  =\sum_{{\cal O}^{(\ell,\ell)}} \lambda_{\phi_1\phi_2{\cal O}} {\cal T}^{\mu_1\ldots \mu_\ell}  {\cal O}^{(\ell,\ell)}_{\mu_1\ldots\mu_\ell}(x_2) + {\rm descendants} \,,
\ee 
where ${\cal T}$ is a tensor structure factor whose explicit form will not be needed. In the latter view, we end up having the commutator of $\hat C$ with ${\cal O}^{(\ell,\ell)}$ which gives the Casimir eigenvalue
\be
[\hat C,{\cal O}^{(\ell,\ell)}(x)] = E_\ell^0  {\cal O}^{(\ell,\ell)}(x)
\ee
where 
\be\label{eq:eigenvalue}
E_{\ell}^p=\Delta\, (\Delta-4)+\ell^2+(2+p)(\ell+\frac{p}{2})
\ee
is the value associated to an operator in the $(\ell+p, \ell)$ or $(\ell, \ell+p)$ Lorentz representations.
Using then eq.(\ref{sch4pt}) one derives a differential equation for each CPW, for any fixed $\Delta$ and $\ell$.

The explicit form of the second order differential operator acting on the CPW or directly on the CB is best derived in the $4+2$-dimensional embedding space. 
The CPW of scalar correlators is parametrized by a single conformal block $G_0^{(0)}(z,\bar z)$.
When acting on scalar operators at $x_1$ and $x_2$, the Lorentz generator can be written as $L_{MN} = L_{1,MN}+L_{2,MN}$, where
\be
L_{i\,MN}=i\Big(X_{i\,M}\frac{\partial}{\partial {X_{i}^N}}-X_{i\,N}\frac{\partial}{\partial {X_{i}^M}}\Big)\,.
\label{LiMN}
\ee
Plugging eq.(\ref{LiMN}) in eq.(\ref{eq:casimir_operator}), one finds after a bit of algebra the Casimir equation \cite{Dolan:2003hv} 
\be
\Delta_{2}^{(a,b;0)} G_0^{(0)}(z,\bar z) = \frac 12 E_{\ell}^0 G_0^{(0)}(z,\bar z)\,,
\label{CasDiffEqp00}
\ee
where $a$ and $b$ are defined in eq.(\ref{abDef}), $u = z\bar z$ and $v=(1-z)(1-\bar z)$. The second-order differential operator $\Delta$ is defined as 
 \begin{equation}
\Delta_{\epsilon}^{(a,b;c)}= D^{(a,b;c)}_z+D^{(a,b;c)}_{\bar z}+\epsilon\,\frac{z\bar z}{z -\bar z}\Big((1-z)\partial_z-(1-\bar z)\partial_{\bar z}\Big)\,,
\label{DeltaGen}
\end{equation}
in terms of the second-order holomorphic operator
 \begin{equation}\label{Dabc}
D^{(a,b;c)}_z\equiv z^2(1-z)\partial_z^2-\big((a+b+1)z^2-c z\big)\partial_z-a b z\,.
\end{equation}
The above derivation can be generalized  for CPWs entering 4-point correlators of tensor fields.
As we have seen in section \ref{sec:seed}, in the most general case the exchange of a given field ${\cal O}^{(\ell,\bar \ell)}$ is not parametrized by a single CPW, but by a set of CPWs
$W^{(i,j)}$, whose number depends on the number of  tensor structures defining the three-point functions $(12{\cal O})$ and $(34\overline{\cal O})$.
In order to derive the second order differential equation satisfied by $W^{(i,j)}$ one has to properly identify the OPE coefficients 
$\lambda^i$ appearing in the generalization of eq.(\ref{phiOPE}) with those in eq.(\ref{sch4pt}).
This is not needed for the seed correlators (\ref{our seed corr}) since the CPW is unique, like in the scalar correlator. For each $p$, we have
\be\label{eq:eigenvalue_problem}
C\, W^{seed}(p) = E_{\ell}^p\, W^{seed}(p),
\ee
where $C$ is the explicit differential form of the Casimir operator to be determined and $E_\ell^p$ is as in eq.(\ref{eq:eigenvalue}). An identical equation is satisfied by $\overline W^{seed}(p)$.
Contrary to the scalar case, the single differential equation (\ref{eq:eigenvalue_problem}) for $W^{seed}(p)$ turns into a system of equations for the 
$p+1$ CBs $G_e^{(p)}$. Let us see how this system of equations can be derived for any $p$.

The action of the Lorentz generators $L_{i,MN}$ on tensor fields should include, in addition to the orbital contribution (\ref{LiMN}), the spin part. 
Recall that $SO(2,4)\simeq SU(2,2)$ and at the level of representations $\bf{8}_{spin}\simeq\bf{4}+\bf{\bar 4}$, where $\bf{4}$ and $\bf{\bar 4}$ represent twistor indices. 
Denoting  by $[\Sigma_{MN}]_a^{\;\;b}$ and $[\overline{\Sigma}_{MN}]^{a}_{\;\;b}$ the generators of $SU(2,2)$ fundamental/anti-fundamental (twistor) representations (see Appendix A of ref.\cite{Elkhidir:2014woa} for details and our conventions),
one can label the 6D spin representations by two integers $(s,\,\bar s)$ which count the number of twistor indices in the $\bf{4}$ and $\bf{\bar 4}$ representations respectively. 
The Lorentz generators acting on generic 6D fields in the $(s,\,\bar s)$ representation are then given by 
\bea
[L_{i\,MN}]_{a_1..\,a_{\bar s} ;\; d_1..\,d_s}^{b_1..\,b_s ;\; c_1..\,c_{\bar s}}
& = & i(X_{i\,M}\partial_{i\,N}-X_{i\,N}\partial_{i\,M})(\delta_{a_1}^{c_1}..\,\delta_{a_{\bar s}}^{c_{\bar s}})(\delta_{d_1}^{b_1}..\,\delta_{d_{s}}^{b_{s}})\nn \\ 
&+&i \bigg([\Sigma_{MN}]_{a_1}^{c_1}\delta_{a_2}^{c_2}..\delta_{a_{\bar s}}^{c_{\bar s}}+[\Sigma_{MN}]_{a_2}^{c_2}\delta_{a_1}^{c_1}..\delta_{a_{\bar s}}^{c_{\bar s}}+..\bigg)\delta_{d_1}^{b_1}..\delta_{d_{s}}^{b_{s}} 
\label{eq:Lorentz_generators}  \\
&+&i\bigg([\overline{\Sigma}_{MN}]_{d_1}^{b_1}\delta_{d_2}^{b_2}..\delta_{d_{s}}^{b_{s}}+[\overline{\Sigma}_{MN}]_{d_2}^{b_2}\delta_{d_1}^{b_1}..\delta_{d_{s}}^{b_{s}}+..\bigg)\delta_{a_1}^{c_1}..\delta_{a_{\bar s}}^{c_{\bar s}}\,. \nn
\eea
We can get rid of all the twistor indices by defining the index-free Lorentz generators
\be\label{eq:lorentz_gen_indexfree}
L_{iMN}=i(X_{iM}\partial_{iN}-X_{iN}\partial_{iM})+i(S_i \overline{\Sigma}_{MN} \partial_{S_i})
+i(\bar S_i \Sigma_{MN} \partial_{\bar S_i}).
\ee
Given any 6D tensor $O(X,S,\bar S)$ , we have
\be
[\hat L_{MN}, O_i(X_i,S_i,\bar S_i)] =  L_{iMN} O_i(X_i,S_i,\bar S_i)\,,
\label{actLhilbert}
\ee
where $\hat L_{MN}$ satisfy the Lorentz algebra
\be 
\label{eq:Lorentz_algebra}
[\hat L_{MN},\hat L_{RS}]=i\Big(\eta_{MS} \hat L_{NR} + \eta_{NR}\hat L_{MS} - \eta_{MR} \hat L_{NS} - \eta_{NS} \hat L_{MR}\Big).
\ee
The explicit form of the Casimir differential operator entering eq.(\ref{eq:eigenvalue_problem}) is obtained by plugging eq.(\ref{eq:lorentz_gen_indexfree}) in eq.(\ref{eq:casimir_operator}).
The single equation (\ref{eq:eigenvalue_problem})  for the CPW turns into a system of second-order coupled differential equations for the $p+1$ conformal blocks $G_e^{(p)}$, $e=0,\ldots,p$, since
the coefficients multiplying the $p+1$ tensor structures in eq.(\ref{eq:CPW_parametrization}) should vanish independently. Schematically
\be
(C-E_\ell^p) \Big(\mathcal K_4 \sum_{e=0}^p G^{(p)}_e (U,V) I_{42}^e J_{42,31}^{p-e}\Big) = \mathcal K_4 \sum_{e=0}^{p} Cas_e^{(p)}(G) I_{42}^e J_{42,31}^{p-e} = 0 \ \  \Rightarrow Cas_e^{(p)}(G) = 0\,,
\label{CasSystem}
\ee
where $Cas_e^{(p)}(G)$ are the $p+1$ Casimir equations, in general each one involving all conformal blocks $G_e^{(p)}$. Determining the Casimir system $Cas_e^{(p)}(G)$ 
is conceptually straightforward but technically involved. The main complication arises from the spin part of the Lorentz generator (\ref{eq:lorentz_gen_indexfree}) that generates products of $SU(2,2)$ invariants not present in eq.(\ref{eq:CPW_parametrization}). The new invariants are linearly dependent and must be eliminated using relations among them. See Appendix A of ref.\cite{Echeverri:2015rwa} for a list of such relations.
This is a lengthy step, that however can be automatized in a computer.
When redundant structures have been eliminated, one is finally able to read from eq.(\ref{CasSystem}) the Casimir system $Cas_e^{(p)}(G)$.
Despite the complicacy of the computation, 
the final system of $p+1$ equations can be written into the following remarkably compact form:
\begin{equation}\label{eq:CasimirSystem}
Cas_e^{(p)}(G)= \Big(\Delta_{2+p}^{(a_e,b_e;c_e)}-\frac{1}{2}\,\big(E_{\ell}^{p}-\varepsilon_{e}^{p}\big)\Big)\,G_e^{(p)}
+A_{e}^{p}\,z\bar z\, L(a_{e-1})\,G_{e-1}^{(p)}+B_{e}\ L(b_{e+1})\,G_{e+1}^{(p)}=0\,,
\end{equation}
where  $e=0,\ldots,p$,
\begin{equation}\label{epsilon_ep}
\varepsilon_{e}^p\equiv\tfrac{3}{4}\,p^2-(1+2e)\,p+2e\,(2+e),\ \ \;A_{e}^p\equiv 2(p-e+1),\ \ \ \;\;B_{e}\equiv \frac{e+1}{2}\,,
\end{equation}
and the coefficients $E_{\ell}^{p}$ are given in eq.(\ref{eq:eigenvalue}). In eq.(\ref{eq:CasimirSystem}) it is understood that $G_{-1}^{(p)}=G_{p+1}^{(p)}=0$. 
An identical system of equations is satisfied by the conjugate CBs $\overline G_e^{(p)}$.
Interestingly enough, only two differential operators enter into the Casimir system: the second-order operator (\ref{DeltaGen}) that already features $p=0$, 
with coefficients $a_e$, $b_e$ and $c_e$ given by 
\begin{equation}
\label{abcExp}
a_e\equiv a,\;\;b_e\equiv b+(p-e),\;\;
c_e\equiv p-e\,,
\end{equation} 
and the new linear operator $L(\mu)$ given by 
\begin{equation}
L(\mu)\equiv -\frac{1}{z-\bar z}\;\Big(\,   z(1-z)\partial_z-\bar z(1-\bar z)\partial_{\bar z}   \,\Big)+\mu.
\label{linearOp}
\end{equation}
Another remarkable property of the Casimir system (\ref{eq:CasimirSystem}) is that, for each given $e$ and $p$, at most three conformal blocks mix with each other in a  sort of ``nearest-neighbour interaction":  
$G_{e}$ mixes only with $G_{e+1}$ and $G_{e-1}$. The Casimir equations at the ``boundaries" $Cas_0^{(p)}$ and $Cas_p^{(p)}$ involve just two blocks.
For $p=0$, the second and third terms in eq.(\ref{eq:CasimirSystem}) vanish and the system trivially reduces to the single equation (\ref{CasDiffEqp00}).

Finding the solution of the system (\ref{eq:CasimirSystem}) is a complicated task, that we address in the next sections.

\section{Shadow Formalism}
\label{sec:shadow}

Another method to obtain CBs in closed analytical form uses the so called shadow formalism.  It was first introduced by Ferrara, Gatto, Grillo, and Parisi \cite{Ferrara:1972xe,Ferrara:1972uq,Ferrara:1972ay,Ferrara:1973vz} and used in ref.\cite{Dolan:2000ut} to get closed form expressions for the scalar CBs. In this section we apply the shadow formalism, using the recent formulation given in ref.\cite{SimmonsDuffin:2012uy}, to get compact expressions for $W^{seed}(p)$ and $\overline W^{seed}(p)$ 
in an integral form for any $p$ and $\ell$.\footnote{The shadow formalism given in an index-free 6D embedding twistor space has also been used in refs.\cite{Fitzpatrick:2014oza,Khandker:2014mpa} to compute CBs in supersymmetric CFTs.}  Using these expressions, we compute the CBs $G_e^{(p)}$ and $\overline{G}_e^{(p)}$ for $\ell=0$ and generic $p$. 
We then provide a practical way to obtain $G_e^{(p)}$ and $\overline{G}_e^{(p)}$ for any $\ell$ in a compact form. 
We finally use this method to compute $G_e^{(p)}$ and $\overline{G}_e^{(p)}$ for $p=1$ and $G_e^{(p)}$ for $p=2$ explicitly.

Despite the power of the above technique, it is computationally challenging to go beyond the $p=2$ case. Moreover, as we will see, 
we do not have any control on the final analytic form of CBs.  
In light of this, we will provide the full analytic solution for $G_e^{(p)}$ and $\overline{G}_e^{(p)}$, for any $p$, only in section \ref{sec:FinalSol}, where we solve directly the set of Casimir differential equations by using an educated ansatz for the solution. The results obtained in this section are however of essential help to argue the proper ansatz.
They will also allow us to get the correct physical asymptotic behaviour of $G_e^{(p)}$ and $\overline{G}_e^{(p)}$ that will  be used as boundary conditions to solve the Casimir system of equations (\ref{eq:CasimirSystem}).
Finally, the explicit computation of $G_e^{(p)}$ and $\overline{G}_e^{(p)}$ for $p=1$ and $G_e^{(p)}$ for $p=2$ using the shadow formalism provides an important consistency check for the validity
of the full general solution (\ref{eq:Ansatz}) to be found in section \ref{sec:FinalSol}.

\subsection{CPW in Shadow Formalism}
\label{subsec:CPWShadow}

We start by briefly reviewing the shadow formalism along the lines of  ref.\cite{SimmonsDuffin:2012uy}, where the reader can find more details. The CPW associated to the exchange of a given operator 
$O_r$ with spin $(\ell,\bar \ell)$ in a correlator of four operators $O_n(X_n)$, $n=1,2,3,4$ (in embedding space and twistor language) is given by 
\be
W_{O^{(\ell,\bar \ell)}}^{(i,j)}(X_i) ={\nu} \! \int \! d^4X_0  \langle  O_1(X_1) O_2(X_2) O_r(X_0,S,\bar S)\rangle_i \overleftrightarrow \Pi_{\ell,\bar \ell} 
 \langle  \widetilde O_{r}(X_0,T,\bar T) O_3(X_3) O_4(X_4)\rangle_j\Big|_M \,,
 \label{shadow}
\ee
where $\nu$ is a normalization factor, the projector gluing two 3-point functions is given by
\be
\overleftrightarrow \Pi_{\ell,\bar \ell}=(\overleftarrow \partial_S X_0 \overrightarrow \partial_T)^\ell (\overleftarrow \partial_{\bar S} \overline{X}_0 \overrightarrow \partial_{\bar T})^{\bar \ell}\,,
\ee
and $\widetilde O_r$ is the shadow operator
\be
\widetilde O_r(X,S,\bar S)\equiv \int d^4 Y \frac{1}{(-2X\cdot Y)^{4-\Delta+\ell+\bar\ell} } O_{\bar r}(Y, Y \bar S, \bar Y S)\,.
\ee
In eq.(\ref{shadow}) we have omitted for simplicity the dependence of $O_n$ on their auxiliary twistors $S_n$, $\bar S_n$, and
the subscripts $i$ and $j$ in $\langle O_1O_2 O_r \rangle$ and $\langle\widetilde O_r O_3 O_4  \rangle$
denote the three point functions stripped of their OPE coefficients:
\be
\langle O_1O_2 O_3\rangle \equiv \sum_i \lambda_{O_1O_2O_3}^i  \langle O_1O_2 O_3\rangle_i \,.
\ee
The integral in eq.(\ref{shadow}) would actually determine the CPW associated to the operator $O_r(X,S,\bar S)$ plus its unwanted shadow counterpart, that corresponds to the exchange of a similar operator but with the
scaling dimension $\Delta\rightarrow 4-\Delta$. The two contributions can be distinguished by their different behaviour under the monodromy transformation $ X_{12}\rightarrow e^{4\pi i}X_{12}$. 
In particular, the physical CPW should transform with the phase $e^{2i\pi(\Delta-\Delta_1-\Delta_2)}$, independently of the Lorentz quantum numbers of the external and exchanged operators. 
This projection on the correct monodromy component explains the subscript $M$ in the bar at the end of eq.(\ref{shadow}).

We use eq.(\ref{shadow}) to get an integral form of $W^{seed}(p)$ and $\overline W^{seed}(p)$ in eq.(\ref{eq:CPW_parametrization}).
The explicit expressions of the needed 3-point functions are given by
\bea
\langle \Phi_1(X_1) F_2(X_2) O^{(\ell,\ell+p)}(X_0)\rangle  &=&  \mathcal K_3(\tau_1,\tau_2,\tau)I_{02}^p J_{0,12}^\ell \,,\nn \\
\langle \Phi_1(X_1) F_2(X_2) \overline O^{(\ell+p,\ell)}(X_0)\rangle  &=&  \mathcal K_3(\tau_1,\tau_2,\tau)K_{1,02}^p J_{0,12}^\ell \,,
\label{3ptWseed}
\eea
where
\be\label{eq:kinematicfactor1}
\mathcal{K}_3(\tau_1,\tau_2,\tau_3)=X_{12}^{\frac{\tau_3-\tau_1-\tau_2}2} X_{13}^{\frac{\tau_2-\tau_1-\tau_3}2}  X_{23}^{\frac{\tau_1-\tau_2-\tau_3}2} ,
\ee
is a kinematic factor and
\be
\label{eq:invar2}
K_{i,jk}            \equiv  \sqrt{\frac{X_{jk}}{X_{ij}X_{ik}}}S_j \overline{\mathbf{X}}_i S_k \,, \ \ \ \ 
\overline{K}_{i,jk} \equiv  \sqrt{\frac{X_{jk}}{X_{ij}X_{ik}}} \bar S_j \mathbf{X}_i \bar S_k \,, \ \ \ 
J_{i,jk}           \equiv  \frac{1}{X_{jk}} \bar S_i  \mathbf{X}_j \overline{\mathbf{X}}_k S_i 
\ee
are $SU(2,2)$ invariants for three-point functions. The ``shadow" 3-point function counterparts are given by 
 \bea
\langle \widetilde O^{(\ell,\ell+p)}(X_0) \Phi_3(X_3) \bar F_4(X_4)\rangle \propto \langle O^{(\ell,\ell+p)}(X_0) \Phi_3(X_3) \bar F_4(X_4)\rangle\Big|_{\Delta\rightarrow 4-\Delta} &=&
 \mathcal K_3\Big|_{\Delta\rightarrow 4-\Delta}\overline K_{3,04}^p J_{0,34}^\ell, \nn \\
\langle \widetilde{\overline O}^{(\ell+p,\ell)}(X_0) \Phi_3(X_3) \bar F_4(X_4)\rangle  \propto  \langle \overline O^{(\ell+p,\ell)}(X_0) \Phi_3(X_3) \bar F_4(X_4)\rangle\Big|_{\Delta\rightarrow 4-\Delta}
 &=& \mathcal K_3\Big|_{\Delta\rightarrow 4-\Delta}I_{40}^p J_{0,34}^\ell. \nn
\eea
Using the above relations, after a bit of algebra, one  can write
\bea
 W^{seed}{(p)}& = & \frac{\nu}{X_{12}^{a_{12}+\frac{\ell}{2}}X_{34}^{a_{34}+\frac{\ell+p}{2}}}\int D^4 X_0 
 \frac{\mathcal N_\ell{(p)}}{X_{01}^{a_{01}+\frac{\ell}{2}}X_{02}^{a_{02}+\frac{\ell+p}{2}}X_{03}^{a_{03}+\frac{\ell+p}{2}}X_{04}^{a_{04}+\frac{\ell}{2}} }\Big|_{M=1} \,,
\label{ShadowIntegral} \\
 \overline W^{seed}{(p)}& = &  \frac{\overline \nu}{X_{12}^{a_{12}+\frac{\ell+p}{2}}X_{34}^{a_{34}+\frac{\ell}{2}}}\int D^4 X_0 
 \frac{\overline{\mathcal N}_\ell{(p)}}{X_{01}^{a_{01}+\frac{\ell+p}{2}}X_{02}^{a_{02}+\frac{\ell}{2}}X_{03}^{a_{03}+\frac{\ell}{2}}X_{04}^{a_{04}+\frac{\ell+p}{2}} }\Big|_{M=1}\,,
\label{ShadowIntegralDual}
\eea
where
\begin{align}
a_{01} &=\frac{\Delta}2+\frac{p}{4}-a,   &  a_{02}  &=\frac\Delta 2-\frac{p}{4}+a, & a_{12} &=\frac{\Delta_1+\Delta_2}2-\frac{\Delta}{2}, \nn  \\  
 a_{03} &=\frac{4-\Delta}2+\frac{p}{4}+b,  & a_{04} &=\frac{4-\Delta}2-\frac{p}{4}-b,  & a_{34} &=\frac{\Delta_3+\Delta_4}2-\frac{4-\Delta}{2}\,,
\end{align}
and
\bea
\mathcal N_\ell{(p)} &\equiv (\bar S S_2)^p (\bar S X_2 \bar X_1 S)^\ell \overleftrightarrow \Pi_{\ell,\ell+p} (\bar S_4 X_3 \bar T)^p (\bar T X_4 \bar X_3 T)^\ell, \label{eq:numerator} \\
\overline{\mathcal N}_\ell{(p)} &\equiv (\bar S_4 S)^p (\bar S X_3 \bar X_4 S)^\ell \overleftrightarrow \Pi_{\ell+p,\ell} ( S_2 \overline X_1 T)^p (\bar T X_1 \bar X_2 T)^\ell.
\label{eq:numerator2}
\eea
We will not need to determine the normalization factors $\nu$ and $\bar \nu$ in eqs.(\ref{ShadowIntegral}) and (\ref{ShadowIntegralDual}). Notice that the correct behaviour of the seed CPWs under 
$X_{12}\rightarrow e^{4\pi i}X_{12}$ is saturated by the factor $X_{12}$ multiplying the integrals in eqs.(\ref{ShadowIntegral}) and (\ref{ShadowIntegralDual}). 
Hence the latter should be projected to their trivial monodromy components $M=1$, as indicated.
Notice that  eqs.(\ref{eq:numerator}) and (\ref{eq:numerator2}) are related by a simple transformation:
\be\label{eq:numVSdual}
\overline{\mathcal N}_\ell{(p)}=\mathcal{P} \mathcal N_\ell{(p)}\Big|_{1\leftrightarrow 3,\;2\leftrightarrow 4},
\ee
where $\mathcal{P}$ is the parity operator. 

We can recast the expression (\ref{eq:numerator}) in a compact and convenient form using some manipulations.
We first define 3 variables
\be
s\equiv X_{12}X_{34} \prod_{n=1}^4 X_{0n},\;
t\equiv\frac{1}{2\sqrt s} \Big(X_{02}X_{03}X_{14}-X_{01}X_{03}X_{24}-(3\leftrightarrow 4) \Big),\;
u\equiv\frac{X_{02}X_{03}X_{34}}{\sqrt s}.
\label{stuDef}
\ee
Then we look for a relation expressing the generic $\mathcal N_{\ell}{(p)}$ in terms of the 
known $\mathcal N{_\ell(0)}$:
\be\label{eq:elegantRelations}
\mathcal N_\ell{(0)} = (-1)^\ell (\ell!)^4\;s^{\ell/2} C_\ell^1(t)\,,
\ee
where $C_\ell^p$ are Gegenbauer polynomials of rank $p$. 
Starting from eq.(\ref{eq:numerator}), after acting with the $S$ and $T$ derivatives, one gets
\be
 {\cal N}_\ell(p) = (\ell!)^2 (\overrightarrow \partial_{\bar S} \overline{X}_0 \overrightarrow \partial_{\bar T})^{\ell +p} \left( (\bar S S_2)^p(\bar S_4 X_3 \bar T)^p (\bar S \Omega \bar T)^\ell\right) \,,
\ee
where we have defined $\,\Omega_{ab}=(X_2 \bar X_1 X_0 \bar X_3 X_4)_{ab}$ . 
In order to relate ${\cal N}_\ell(p)$ above to ${\cal N}_{\ell+p}(0)$ in eq.(\ref{eq:elegantRelations}),
we look for an operator $\widetilde{\mathcal D}$ satisfying
\be
\widetilde{\mathcal D}^p \,(\overrightarrow \partial_{\bar S} \overline{X}_0 \overrightarrow \partial_{\bar T})^{\ell+p} (\bar S \Omega \bar T)^{\ell+p}=  (\overrightarrow \partial_{\bar S} \overline{X}_0 \overrightarrow \partial_{\bar T})^{\ell +p} \left( (\bar S S_2)^p(\bar S_4 X_3 \bar T)^p (\bar S \Omega \bar T)^\ell\right)\,.
\label{ShadowOpDef} 
\ee
We deduce that  $\widetilde{\mathcal D}$ should be bilinear in $\bar S_4$ and $S_2$ and should commute with $(\overrightarrow \partial_{\bar S} \overline{X}_0 \overrightarrow \partial_{\bar T})$.
In addition to that, it should have the correct scaling in $X$'s and should be gauge invariant, namely it should be well defined on the light-cone $X^2=0$ and preserve the conditions (\ref{gaugered}).
It is not difficult to see that the choice $\widetilde{\mathcal D}= \mathcal D/(8X_{01}X_{04})$, where
\be
\mathcal D= (\bar S_4 X_{0} \bar \Sigma^N S_2)\frac{\partial}{\partial X_{2}^N} 
\ee
fulfills all the requirements.
One has $\widetilde{\mathcal D}  (\bar S \Omega \bar T) = (\bar S S_2)(\bar S_4 X_3 \bar T)$. Iterating it $p$ times gives the desired relation:
\be\label{eq:numeratorGeneralP}
\mathcal N{_\ell(p)}\propto \widetilde{\mathcal{D}}^p \mathcal N_{\ell+p}{(0)}\,. 
\ee
The operator ${\mathcal D}$ annihilates all the scalar products with the exception of $X_{12}$, in which case we have
$\mathcal{D}X_{12}=I_2$, and we define 
\be
I_1\equiv X_{03}\,J_{42,30}, \ \ \ \ \ I_2\equiv X_{01}\,J_{42,01}\,.
\ee
The action on the $s,\,t,$ and $u$ variables is
\be 
\mathcal{D}\,s=X_{12}^{-1}\,s\,I_2,\;\;
\mathcal{D}\,t=-\frac{1}{2}\, X_{12}^{-1}\,(u^{-1}\,I_1+t\,I_2),\;\;
\mathcal{D}\,u^{-1}=\frac{1}{2}\,X_{12}^{-1}\,u^{-1}\,I_2\,,
\ee
on Gegenbauer polynomials is
\be 
\mathcal{D}\,C_{n}^{\lambda}(t)=2\lambda\,C_{n-1}^{\lambda+1}(t)\,\mathcal{D}\,t\,,
\ee
and vanishes on $J_{42,01}$ and $J_{42,30}$. Using recursively the identity for Gegenbauer polynomials 
\be 
\frac{n}{2\lambda}\,C_{n}^{\lambda}(t)-t\,C_{n-1}^{\lambda+1}(t)=-C_{n-2}^{\lambda+1}(t)\,,
\ee
we can write the following expression for $\mathcal N_\ell{(p)}$:
\be\label{eq:numeratorCompact}
\mathcal N_\ell{(p)}\propto
s^{\tfrac{\ell}{2}}\;
\sum_{w=0}^p \binom {p} {w}
u^w\,C_{\ell-w}^{p+1}(t)\;I_1^{p-w}I_2^w,
\ee
where $\binom {p} {w}$ is the binomial coefficient and for compactness we have defined the dimensionful tensor structures
Combining together eqs.(\ref{ShadowIntegral}), (\ref{ShadowIntegralDual}), (\ref{eq:numVSdual}), (\ref{stuDef}) and (\ref{eq:numeratorCompact}) we can finally write 
\bea
W^{seed}{(p)} \! & = & \! \nu'\sum_{w=0}^p \binom {p} {w}\,\frac{1}{X_{12}^{a_{12}+\frac{w}{2}}X_{34}^{a_{34}+\frac{p-w}{2}}}\int \!\! D^4 X_0 
 \frac{C_{\ell-w}^{p+1}(t)\;I_1^{p-w}I_2^w}
 {X_{01}^{a_{01}+\frac{w}{2}}X_{02}^{a_{02}+\frac{p-w}{2}}X_{03}^{a_{03}+\frac{p-w}{2}}X_{04}^{a_{04}+\frac{w}{2}} }\bigg|_{M=1}, \nn \\
\label{eq:MasterRelation}
\overline W^{seed}{(p)} \!&=&\! \bar\nu' \sum_{w=0}^p \binom {p} {w}\,\frac{1}{X_{12}^{a_{12}+\frac{p-w}{2}}X_{34}^{a_{34}+\frac{w}{2}}}\int \!\! D^4 X_0 
 \frac{C_{\ell-w}^{p+1}(t)\;I_1^{w}I_2^{p-w}}
 {X_{01}^{a_{01}+\frac{p-w}{2}}X_{02}^{a_{02}+\frac{w}{2}}X_{03}^{a_{03}+\frac{w}{2}}X_{04}^{a_{04}+\frac{p-w}{2}} }\bigg|_{M=1}
\eea
where $\nu^\prime$ and $\bar \nu^\prime$ are undetermined normalization factors.

\subsection{Seed Conformal Blocks and Their Explicit Form for $\ell=0$}

\label{subsec:Shadowl0}

The computation of the CBs  $ G_e^{(p)}$ and $\overline G_e^{(p)}$ starting form  eq.(\ref{eq:MasterRelation}) is a non-trivial task for generic $\ell$ and $p$, since
we are not aware of a general formula for an integral that involves  $C_{\ell-w}^{p+1}(t)$ for $p\neq 0$. For any given $\ell$, one can however expand the Gegenbauer polynomial, in which case
the CBs $G_e^{(p)}$ and $\overline G_e^{(p)}$ can be computed. In this subsection we 
discuss the structure of CBs for generic $\ell$ and compute $G_e^{(p)}$ and $\overline G_e^{(p)}$ for $\ell=0$ and generic $p$.

Recalling the definition of $t$ in eq.(\ref{stuDef}),  one realizes that the Gegenbauer polynomials in eq.(\ref{eq:MasterRelation}), when expanded, do not give rise to intrinsically new integrals but just amounts to shifting the exponents in the denominator. The tensor structures in the numerators bring $p$ open indices in the form $X_0^{N_1}\ldots X_0^{N_p}$, which can be removed by using eq.(3.21) in ref.~\cite{SimmonsDuffin:2012uy}. In this way the problem is reduced
to the computation of scalar integrals in $2h=2(2+p)$ effective dimensions, of the form: 
\be 
I^{(h)}_{A_{02},\,A_{03},\,A_{04}}\equiv
\int D^{2h} X_0 
 \frac{1}{X_{01}^{A_{01}}X_{02}^{A_{02}}X_{03}^{A_{03}}X_{04}^{A_{04}} }\bigg|_{M=1},
\ee
where $A_{01}+A_{02}+A_{03}+A_{04}=2h$. The capital $A_{0i}$ are used for the exponents in the denomentaor  with all possible shifts introduced by the Gegenbaur polynomials. This integral is given by 
\be\label{eq:ShadowIntegral} 
I^{(h)}_{A_{02},\,A_{03},\,A_{04}}\propto 
X_{13}^{A_{04}-h} X_{14}^{A_{02}+A_{03}-h} X_{24}^{-A_{02}} X_{34}^{h-A_{03}-A_{04}}
\times R^{(h)}(z,\bar z;\,A_{02},A_{03},A_{04}),
\ee
where  
\bea
R^{(h)}(z,\bar z;\,A_{02},A_{03},A_{04}) &\equiv& \Big(-\frac{\partial}{\partial v}\Big)^{h-1}\,f(z;\,A_{02},A_{03},A_{04})f(\bar z;\,A_{02},A_{03},A_{04}),\\
f(z;\,A_{02},A_{03},A_{04})&\equiv& \hspace{0.7mm}_2F_1(A_{02}-h+1,\,-A_{04}+1;\,-A_{03}-A_{04}+h+1;\,z).
\eea
The derivative $-\partial/\partial v$ in $(z,\bar z)$ coordinates equals
\be
-\frac{\partial}{\partial v} = \frac{1}{z-\bar z}\Big(z\frac{\partial}{\partial z}-\bar z\frac{\partial}{\partial \bar z} \Big).
\ee
In the case of $\ell=0$, all the above manipulations simplify drastically. The Gegenbauer polynomials $C_{\ell-w}^{p+1}(t)$ vanishe for all the values $w$ except for $w=0$, leaving only one type of tensor structure: $I_1^p$ for $W^{seed}(p)$ and $I_2^p$ for $\overline W^{seed}(p)$. This leads to a one-to-one correspondence between CBs and integrals:
\bea
G_e^{(p)} &\propto& X_{13}^{p-e}X_{34}^e\,\mathcal{K}_4^{-1}\,
I^{(2+p)}_{a_{02}+\frac{p}{2},\,a_{03}+\frac{p}{2},\,a_{04}+e}\propto (z\bar z)^{\frac{\Delta+\frac{p}{2}}{2}}
R^{(2+p)}(z,\bar z;\,a_{02}+\frac{p}{2},a_{03}+\frac{p}{2},a_{04}+e), \label{Gwpl0} \\
\overline{G}_e^{(p)} &\propto& X_{12}^{e}X_{13}^{p-e}\,\mathcal{K}_4^{-1}\,
I^{(2+p)}_{a_{02}+e,\,a_{03}+p-e,\,a_{04}+\frac{p}{2}}\propto (z\bar z)^{\frac{\Delta-\frac{p}{2}}{2}+e}
R^{(2+p)}(z,\bar z;\,a_{02}+e,a_{03}+p-e,a_{04}+\frac{p}{2}). \nn
\eea
We have omitted here the relative factors between different CBs. They must be restored if one wants to check that $G_e^{(p)}$ and $\overline G_e^{(p)}$ 
in eq.(\ref{Gwpl0}) satisfy the Casimir system (\ref{eq:CasimirSystem}).
For generic $\ell$ the CBs are a sum of expressions like eq.(\ref{Gwpl0}) with different shifts of the parameters $A_{0i}$, weighted by the relative constants and powers of $v$ (coming from the Gegenbauer polynomial). 
Since all these terms have $p+1$ derivatives with respect to $v$, the highest power in $1/(z-\bar z)$ appearing in $G_e^{(p)}$ and $\overline{G}_e^{(p)}$ is
\be \label{overallpower}
\Big(\,\frac{1}{z-\bar z}\,\Big)^{1+2\,p}.
\ee
The asymptotic behaviour of the CBs when $z,\bar z\rightarrow 0$ ($u\rightarrow 0$, $v\rightarrow 1$) for $\ell=0$ is easily obtained from eq.(\ref{Gwpl0}) by  noticing that 
$ R^{(h)}(z,\bar z;\,A_{02},A_{03},A_{04})$ is constant in this limit. Then we have
\be
 \lim_{z\rightarrow 0,\,\bar z\rightarrow 0} G_e^{(p)} \propto (z\bar z)^{\frac{\Delta}2+\frac{p}{4}} \,, \label{zzbar0limit0}  \ \ \ \ \ \ 
\lim_{z\rightarrow 0,\,\bar z\rightarrow 0} \overline G_e^{(p)} \propto (z\bar z)^{\frac{\Delta}2-\frac{p}{4}+e}\,. 
\ee
By knowing that the CBs should be proportional to the factor in eq.(\ref{overallpower}), we can refine eq.(\ref{zzbar0limit0})  and write 
\bea
&& \lim_{z\rightarrow 0,\,\bar z\rightarrow 0} G_e^{(p)} \propto \frac{(z\bar z)^{\frac{\Delta}2+\frac{p}{4}}}{(z-\bar z)^{1+2p}} (z^{1+2p}- \bar z^{1+2p}) \,, \label{zzbar0limit} \\
&& \lim_{z\rightarrow 0,\,\bar z\rightarrow 0} \overline G_e^{(p)} \propto   \frac{(z\bar z)^{\frac{\Delta}2-\frac{p}{4}+e}}{(z-\bar z)^{1+2p}} (z^{1+2p}- \bar z^{1+2p})\,.
\label{zzbar0limitCon}
\eea
Notice that the behavior (\ref{zzbar0limit}) and (\ref{zzbar0limitCon}) of the CBs for $z,\bar z\rightarrow 0$ when $\ell=0$ is not guaranteed to be straightforwardly extended for any $\ell\neq 0$.
Indeed, we see from eq.(\ref{eq:MasterRelation}) that for a given $p$, the generic CPW is obtained when $\ell \geq p$, in which case all terms in the sum over $w$ are present.
All the values of $\ell < p$ should be treated separately.


\subsection{Computing the Conformal Blocks for $\ell\neq 0$}
\label{subsec:CBlneq0}

A useful expression of the CBs for generic values of $\ell$ can be obtained using eq.(\ref{eq:numeratorGeneralP}) and the known closed form of $W^{seed}{(0)}$. Recall that
\be
 W^{seed}(0)  
  =  \left(\frac{X_{14}}{X_{13}}\right)^{b}\left(\frac{X_{24}}{X_{14}}\right)^{-a}\frac{G_0^{(0)}(Z,\bar Z)}{X_{12}^{\frac{\Delta_1+\Delta_2}{2}} X_{34}^{\frac{\Delta_3+\Delta_4}{2}}} \,,
\label{DolanOsbornShadow}
\ee
where $a$ and $b$ are as in eq.(\ref{abDef}) for $p=0$ and $G^{(0)}(z,\bar z)$ are the known scalar CBs \cite{Dolan:2000ut,Dolan:2003hv}
\be
G_0^{(0)}(z,\bar z)=G_0^{(0)}(z,\bar z;\Delta,l,a,b)=(-1)^\ell \frac{z \bar z}{z-\bar z}\left(k_{ \frac{\Delta+\ell}{2}}^{(a,b;0)}(z)k_{\frac{\Delta-\ell-2}{2}}^{(a,b;0)}(\bar z)-(z\leftrightarrow \bar z)\right)\,,
\ee
expressed in terms of the function\footnote{We adopt here the notation first used in ref.\cite{Rattazzi:2008pe} for this function, but notice
the slight difference in the definition: $k_{\rho}^{there}=k_{\rho/2}^{here}$.}
\begin{equation}
\label{krhoDef}
k_\rho^{(a,b;c)}(z)\equiv z^\rho\hspace{1mm}_2F_1(a+\rho,\,b+\rho;\,c+2\rho;\,z) \,.
\end{equation}
Comparing eq.(\ref{DolanOsbornShadow}) with eq.(\ref{eq:MasterRelation}) for $p=0$, one can extract the value of the shadow integral in closed form for generic spin $\ell$ \cite{SimmonsDuffin:2012uy}:
\be\label{eq:IntegralRessumed}
I_{\ell}\equiv
\int D^{4} X_0 
 \frac{C_{\ell}^1(t)}{X_{01}^{a_{01}}X_{02}^{a_{02}}X_{03}^{a_{03}}X_{04}^{a_{04}} }\Big|_{M=1}\propto
\left(\frac{X_{14}}{X_{13}}\right)^{b}\left(\frac{X_{24}}{X_{14}}\right)^{-a} \frac{G_0^{(0)}(Z,\bar Z;\Delta,\ell,a,b)}{X_{12}^{\frac{\Delta}{2}}X_{34}^{\frac{4-\Delta}{2}}}\,.
\ee
Using the relations (\ref{eq:elegantRelations}) and (\ref{eq:numeratorGeneralP}) one can recast $W^{seed}(p)$  and $\overline W^{seed}(p)$  in the form
\bea
W^{seed}(p) & \propto & \frac{\mathcal{D}_{N_1}...\mathcal{D}_{N_p}}{X_{12}^{a_{12}+\frac{\ell}{2}}X_{34}^{a_{34}}}\;X_{12}^{\frac{\ell+p}{2}}\int D^4 X_0  \frac{C_{\ell+p}^1(t)X_{0}^{N_1}...X_{0}^{N_p}}{X_{01}^{a_{01}+\frac{p}{2}}X_{02}^{a_{02}}X_{03}^{a_{03}}X_{04}^{a_{04}+\frac{p}{2}}}\bigg|_{M=1}\,,
\label{FinalShadow} \nn \\
\overline W^{seed}(p)& \propto & \frac{\overline{\mathcal{D}}_{N_1}...\overline{\mathcal{D}}_{N_p}}{X_{12}^{a_{12}}X_{34}^{a_{34}+\frac{\ell}{2}}}\;X_{34}^{\frac{\ell+p}{2}}\int D^4 X_0  \frac{C_{\ell+p}^1(t)X_{0}^{N_1}...X_{0}^{N_p}}{X_{01}^{a_{01}}X_{02}^{a_{02}+\frac{p}{2}}X_{03}^{a_{03}+\frac{p}{2}}X_{04}^{a_{04}}}\bigg|_{M=1}\,,
\label{FinalShadowConjugate}
\eea
where $\overline {\cal D}= \mathcal{P} {\cal D}|_{1\leftrightarrow 3,2\leftrightarrow 4}$, as follows from eq.(\ref{eq:numVSdual}), ${\mathcal D}= {\mathcal D}_M X^{M}_0$, $\overline{\mathcal D}=\overline{ {\mathcal D}}_M X^{M}_0$ .
 The tensor integral is evaluated using $SO(4,2)$ Lorentz symmetry. One writes
\be
\int D^4 X_0  \frac{C_{\ell+p}^1(t)X_{0}^{M_1}...X_{0}^{M_p}}{X_{01}^{a_{01}+\frac{p}{2}}X_{02}^{a_{02}}X_{03}^{a_{03}}X_{04}^{a_{04}+\frac{p}{2}}}=\sum_n A_n(X_i) \,\, \tau_{n}^{M_1...M_p}(X_i) \,,
\label{ShadowOpenIndices}
\ee
where $n$ runs over all possible rank $p$ traceless symmetric tensors $\tau_n$ which can be constructed from $X_1,X_2,X_3,X_4$ and  $\eta_{MN}$'s, with arbitrary scalar coefficients $A_n$ to be determined. 
Performing all possible contractions, which do not change the monodromy of the integrals,  the $A_n$ coefficients can be solved as linear combinations of the scalar block integrals $I_{\ell}$ defined in eq.(\ref{eq:IntegralRessumed}), 
with shifted external dimensions. 

In this way, we have computed the CBs $G_e^{(p)}$ with $p=1,2$  and  $\overline G_e^{(p)}$ with $p=1$  for general $\Delta,\ell,a,b$.  We have also verified that the CBs $\overline G_e^{(1)}$ obtained from  $G_e^{(1)}$ using eqs.(\ref{WbarFromW}) and (\ref{WbarDiff}) agree with those arising from the direct shadow computation.
There is a close connection among  the CBs  $G_e^{(p)}$ and $\overline G_{p-e}^{(p)}$, for any $p$. More on this point in section \ref{sec:FinalSol}.
In all cases the CBs satisfy the Casimir system (\ref{eq:CasimirSystem}). 

As mentioned at the end of subsection \ref{subsec:Shadowl0}, the asymptotic behaviour of the CBs for $z,\bar z\rightarrow 0$ depends on whether $\ell\geq p$ or not. For $p=1$ we can expand the obtained solutions, which for $\ell \geq 1$ read as
\bea
& \lim_{z\rightarrow 0,\,\bar z\rightarrow 0} G_e^{(1)}\propto \frac{(z\bar z)^{\frac{\Delta-\ell}2+\frac{1}{4}}}{(z-\bar z)^{3}} \Big(\bar z^{\ell+e+2} - (z\leftrightarrow \bar z)\Big) \,, \ \ \ \ \ell\geq 1
\label{zzbar0limitp1l} \\
& \lim_{z\rightarrow 0,\,\bar z\rightarrow 0} \overline G_e^{(1)}\propto    \frac{(z\bar z)^{\frac{\Delta-\ell}2-\frac{1}{4}}}{(z-\bar z)^{3}} \Big(z^{e} \bar z^{\ell+3} - (z\leftrightarrow \bar z)\Big) \,, \ \ \ \ \ell \geq 1 \,,
\label{zzbar0limitp1lCon}
\eea
while for $\ell=0$ they match eqs.(\ref{zzbar0limit}) and (\ref{zzbar0limitCon}). The above relations, together with eqs.(\ref{overallpower}), (\ref{zzbar0limit}) and (\ref{zzbar0limitCon}),
will allow us to settle the problem of the boundary values of the CBs for any value of $p$ and $\ell$, that will be reported in eqs.(\ref{zzbar0limitAnypl})
and (\ref{zzbar0limitAnyplConj}). The explicit form of $G_e^{(p)}$ found for $p=2$ using the shadow formalism provides a further check of the whole derivation.


\section{Solving the System of Casimir Equations}
\label{sec:FinalSol}

The goal of this section is to find the explicit form of the conformal blocks $G_e^{(p)}$ and $\overline G_e^{(p)}$ appearing in eq.(\ref{eq:CPW_parametrization}) by solving the Casimir system~(\ref{eq:CasimirSystem}).
In doing it we adopt and expand the methods introduced by Dolan and Osborn in refs.~\cite{Dolan:2003hv,Dolan:2011dv} to obtain 6D scalar conformal blocks. 
We will mostly focus on the blocks $G_e^{(p)}$, since the same analysis will apply to $\overline G_e^{(p)}$ with a few modifications that we will point out.

Before jumping into details let us outline the main logical steps of our derivation. 
We first find, with the guidance of the results obtained in section \ref{sec:shadow},  the behaviour of $G^{(p)}_e$ and $\overline G_e^{(p)}$ in the limit $z,\bar z\rightarrow 0$ 
in which the Casimir system (\ref{eq:CasimirSystem}) can be easily solved.
Using this information and eq.(\ref{overallpower}), we then write an educated ansatz for the form of the CBs. 
Using this ansatz, we reduce the problem of solving a system of linear partial differential equations of second order in two variables to a system of linear algebraic equations for the unknown coefficients entering the ansatz. 
Then we show that the non-zero coefficients in the ansatz admit a geometric interpretation. They form a two-dimensional lattice  with an octagon shape structure.
This interpretation allows us to precisely predict which coefficients enter in our ansatz for any value of $p$. 
Finally, we show that the linear algebraic system admits a recursive solution and we discuss the complexity of deriving full solutions for higher values of $p$.

\subsection{Asymptotic Behaviour} 
\label{subsec:limitz0}

Not all solutions of the Casimir system (\ref{eq:CasimirSystem}) give rise to sensible CBs. The physical CBs are obtained
by demanding the correct boundary values for  $G_e^{(p)}$ and $\overline G_e^{(p)}$. 
Possible boundary values are given by considering the OPE limit 
$z,\bar z\rightarrow 0$ of  $W^{seed}(p)$ and $\overline W^{seed}(p)$.
The limits of $G_e^{(p)}$ and $\overline G_e^{(p)}$ for $z,\bar z\rightarrow 0$ could be computed  by a careful analysis of tensor structures. 
This analysis has been partly  done in section \ref{sec:shadow}, where we have obtained the boundary values of $G_e^{(p)}$ and $\overline G_e^{(p)}$
for $z,\bar z\rightarrow 0$ for special values of $p$ and/or $\ell$. Luckily enough, there will be no need to extend such analysis because the form of the system (\ref{eq:CasimirSystem}) in the OPE limit, together with 
eqs.(\ref{zzbar0limit}), (\ref{zzbar0limitp1l}) and (\ref{zzbar0limitp1lCon}), will clearly indicate the general form of the boundary values of $G_e^{(p)}$ and $\overline G_e^{(p)}$.

Let us then consider the form of the conformal blocks $G_e^{(p)}$ in the limit $z,\bar z\rightarrow 0$, with $z\rightarrow 0$ {\it taken first}.
In this limit
\begin{equation}
G_e^{(p)} \rightarrow  N_e z^{\lambda^{(e)}} \bar z^{\bar \lambda^{(e)}} \,,
\label{Gzz0}
\end{equation}
where $N_e$, $\lambda^{(e)}$ and $\bar \lambda^{(e)}$ are parameters to be determined. For simplicity of notation we have omitted their $p$-dependence.
The differential operators (\ref{DeltaGen}) and (\ref{linearOp}), when acting on eq.(\ref{Gzz0}) give, at leading order in $z$ and $\bar z$, 
\bea
&& \Delta_\epsilon^{(a_e,b_e;c_e)} \rightarrow \lambda^{(e)}(\lambda^{(e)}-1)+c_e (\lambda^{(e)} + \bar \lambda^{(e)})+\bar \lambda^{(e)} (\bar \lambda^{(e)}-1)- \epsilon  \lambda^{(e)}\,, \label{DiffCaszz0}  \\
&& L(\mu)\rightarrow \frac{1}{\bar z} ( \lambda^{(e)}-\bar \lambda^{(e)}) \,.
\label{DiffLzz0}
\eea
Let us now focus on the specific  equation $Cas_e^{(p)}$ with $e=p$.
In the limit $z,\bar z \rightarrow 0$ it reads
\bea
Cas_p^{(p)}(G)\rightarrow && N_p \bigg( \lambda^{(p)}(\lambda^{(p)}-1)+\bar\lambda^{(p)}(\bar\lambda^{(p)}-1)-(p+2) \lambda^{(p)}-\frac 12 (E_{\ell,p}-\epsilon_p^p)\bigg) z^{\lambda^{(p)}}\bar z^{\lambda^{(p)}} \nn \\
&& +2 N_{p-1} (\lambda^{(p-1)} -\bar \lambda^{(p-1)}) 
z^{\lambda^{(p-1)}+1}\bar z^{\bar\lambda^{(p-1)}}= 0 \,.
\label{Caszzb0}
\eea
For generic values of $\ell$, we have $\lambda^{(e)}\neq \bar\lambda^{(e)}$. Hence we cannot have $\lambda^{(p-1)}+1< \lambda^{(p)}$ in eq.(\ref{Caszzb0}), since this would imply
that the last term dominates in the limit and $N_{p-1}$ vanishes, in contradiction with the initial hypothesis (\ref{Gzz0}).

Let us first consider the case in which  $\lambda^{(p-1)}+1> \lambda^{(p)}$, so that the terms in the second row of eq.(\ref{Caszzb0}), coming from $G_{p-1}^{(p)}$, vanish.
It is immediate to see that the only sensible solution for $\lambda^{(p)}$ and $\bar \lambda^{(p)}$ that reproduce the known OPE limit for the $p=0$ case is
\be
\lambda^{(p)} =  \frac{\Delta-\ell}2+\frac{p}{4}, \ \ \ \ \ \  \bar \lambda^{(p)} =   \frac{\Delta+\ell}2+\frac p4\,. 
 \label{alpha2Ex1}  
 \ee
Notice that eq.(\ref{alpha2Ex1}) agrees with the asymptotic behaviour for the CBs $G_e^{(p)}$ found in eq.(\ref{zzbar0limitp1l}) for $e=p=1$ and $\ell\geq 1$. 
Consider now the equation  $Cas_{p-1}^{(p)}$. For $z,\bar z \rightarrow 0$ we have
\bea
Cas_{p-1}^{(p)}(G)\rightarrow && N_{p-1} \bigg(\lambda^{(p-1)}(\lambda^{(p-1)}-1)+\bar\lambda^{(p-1)}(\bar\lambda^{(p-1)}-1)+(\lambda^{(p-1)}+\bar \lambda^{(p-1)})-(p+2) \lambda^{(p-1)}
 \nn \\
&& -\frac 12 (E_{\ell,p}-\epsilon_{p-1}^p)\bigg)  z^{\lambda^{(p-1)}}\bar z^{\bar\lambda^{(p-1)}}+\frac p2 N_p(\lambda^{(p)} - \bar \lambda^{(p)})  z^{\lambda^{(p)}}\bar z^{\bar \lambda^{(p)}-1}   \nn \\
&& + 4 N_{p-2} (\lambda^{(p-2)} - \bar \lambda^{(p-2)}) z^{\lambda^{(p-2)}+1}\bar z^{\bar\lambda^{(p-2)}}= 0 \,.
\label{Caszzb1}
\eea
According to eq.(\ref{zzbar0limitp1l}), we expect  $\lambda^{(p-2)}=\lambda^{(p-1)}=\lambda^{(p)}$, $\bar \lambda^{p-1}=\bar \lambda^{(p)}-1$,  $\bar \lambda^{p-2}=\bar \lambda^{(p)}-2$ 
in eq.(\ref{Caszzb1}).  In this case the last term is higher order in $z$ and eq.(\ref{Caszzb1}) is satisfied by simply taking
\be
\frac{N_{p-1}}{N_p} = -\frac{\ell p}{2(\ell+p)}\,.
\ee
Notice that we have tacitly assumed above that $\lambda^{(p)}-\bar \lambda^{(p)}=-\ell$ does not vanish, i.e. $\ell\neq 0$.
For $\ell=0$, more care is required and one should consider the first subleading term in $\bar z$ in the expansion (\ref{Gzz0}).

The above analysis can be iteratively repeated until the last equation $Cas_0^{(p)}$ is reached and all the coefficients $N_e$, $\lambda^{(e)}$ and $\bar \lambda^{(e)}$ are determined.
Analogously to the $\ell=0$ case in eq.(\ref{Caszzb1}), all the low spin cases up to $\ell=p$ should be treated separately at some step in the iteration, as already pointed out in subsection 
\ref{subsec:Shadowl0}.
Skipping the detailed derivation, the final values of $\lambda^{(e)}$ and $\bar \lambda^{(e)}$ are given by
\bea
&& \lambda^{(e)} = \lambda^{(p)} \,, \hspace{2cm} \ \ \  \forall \ell=0,1,2,\ldots \nn \\
&&   \bar\lambda^{(e)}=  \bar\lambda^{(p)} - (p-e)  \,, \ \ \ \ \ \  \forall \ell=p-e,p-e+1,\ldots \nn \\
&&   \bar\lambda^{(e)} =  \bar\lambda^{(p)}   \,, \hspace{2cm} \ \ \  \forall \ell=0,1,\ldots, p-e-1 \,,
\eea
where $\lambda^{(p)}$ and $\bar \lambda^{(p)}$ are as in eq.(\ref{alpha2Ex1}) and $e=0,\ldots,p-1$. 
The asymptotic behaviour of the CBs in the OPE limit is given for any $\ell$ and $p$ by
\be
\lim_{z\rightarrow 0,\,\bar z\rightarrow 0} G_e^{(p)}\propto \frac{(z\bar z)^{\lambda^{(p)}}}{(z-\bar z)^{1+2p}} \Big(\bar z^{\bar\lambda^{(e)}-\lambda^{(p)}+1+2p} - (z\leftrightarrow \bar z)\Big) \,.
\label{zzbar0limitAnypl}
\ee
We do not report the explicit form of the normalization factors $N_e$, since they will be of no use in what follows.

We still have to consider the case in which $\lambda^{(p-1)}+1= \lambda^{(p)}$ in eq.(\ref{Caszzb0}). By looking at eq.(\ref{zzbar0limitp1lCon}), it is clear that  this case
corresponds to the asymptotic behaviour of the conjugate CBs $\overline G_e^{(p)}$. We do not report here the similar derivation of the Casimir equations for $\overline G_e^{(p)}$ in the OPE limit.
It suffices to say that the analysis closely follows the ones made for $G_e^{(p)}$ starting now from the equation with $e=0$.
If we denote by 
\begin{equation}
\overline G_e^{(p)} \rightarrow  \bar N_e z^{\omega^{(e)}} \bar z^{\bar \omega^{(e)}}
\label{Gbarzz0}
\end{equation}
the boundary behaviour of $\overline G_e^{(p)}$  when $z,\bar z\rightarrow 0$ ($z\rightarrow 0$ taken first), one finds
\bea
&&   \omega^{(e)}=  \omega^{(0)} +e  \,, \hspace{1.5cm} \forall \ell=0,1,2,\ldots \nn \\
&& \bar\omega^{(e)} = \bar\omega^{(0)} \,, \hspace{2.2cm}  \forall   \ell=p-e,p-e+1,\ldots \\
&&   \bar\omega^{(e)} =  \bar\omega^{(0)}+e   \,, \hspace{1.2cm} \ \ \  \forall \ell=0,1,\ldots, p-e-1 \nn
\eea
where
\be
\omega^{(0)} =  \frac{\Delta-\ell}2-\frac{p}{4}, \ \ \ \ \ \  \bar \omega^{(0)} =   \frac{\Delta+\ell}2-\frac p4\,. 
 \label{alpha2Ex0}  
 \ee
The asymptotic behaviour of the conjugate CBs are given for any $\ell$ and $p$ by
\be
\lim_{z\rightarrow 0,\,\bar z\rightarrow 0} \overline G_e^{(p)}\propto \frac{(z\bar z)^{\omega^{(e)}}}{(z-\bar z)^{1+2p}} \Big( \bar z^{\bar\omega^{(e)}-\omega^{(e)}+1+2p} - (z\leftrightarrow \bar z)\Big) \,.
\label{zzbar0limitAnyplConj}
\ee

\subsection{The Ansatz}

The key ingredient of the ansatz is the function $k_\rho^{(a,b;c)}(z)$ defined in eq.(\ref{krhoDef}), which is an eigenfunction of the hyper-geometric like operator $D^{(a,b;c)}_z$:
\begin{equation}
\label{krhoEigen}
D^{(a,b;c)}_z \;k_\rho^{(a,b;c)}(z)=\rho\,(\rho+c-1)\;k_\rho^{(a,b;c)}(z).
\end{equation}
Using eq.(\ref{krhoEigen}) one can define an eigenfunction of the operator $\Delta_{0}^{(a,b;c)}$ as the product of two $k$'s:
\bea
\mathcal{F}_{\rho_1,\,\rho_2}^{(a,b;c)}(z,\bar z)      &\equiv & k_{\rho_1}^{(a,b;c)}(z)k_{\rho_2}^{(a,b;c)}(\bar z),\\
\mathcal{F}_{\rho_1,\,\rho_2}^{\pm\;(a,b;c)}(z,\bar z) &\equiv & \mathcal{F}_{\rho_1,\,\rho_2}^{(a,b;c)}(z,\bar z) \pm \mathcal{F}_{\rho_1,\,\rho_2}^{(a,b;c)}(\bar z,z).
\eea
These functions played an important role in ref.\cite{Dolan:2003hv} for the derivation of an analytic closed expression of the scalar CBs in even space-time dimensions.
In our case, the situation is much more complicated, because we have different blocks appearing in the Casimir equations.
We notice, however, that the second order operator $\Delta$ in each equation $Cas_e^{(p)}$ acts only on the block $G_e^{(p)}$, while the blocks
$G_{e-1}^{(p)}$ and $G_{e+1}^{(p)}$ are multiplied by first order operators only. 
Since, as we will shortly see, first order derivatives and factors of $z$ and $\bar z$ acting on the functions ${\cal F}$ 
can always be expressed in terms of functions ${\cal F}$ with shifted parameters, a reasonable ansatz for the CBs is to take each $G_e$ proportional to  a sum
of functions of the kind $\mathcal{F}_{\rho_1,\,\rho_2}^{(a_e,b_e;c_e)}(z,\bar z)$ for some $\rho_1$ and $\rho_2$.
Taking also into account eq.(\ref{overallpower}), found using the shadow formalism, 
the form of the ansatz for the blocks $G^{(p)}_e$ should be\footnote{Recall that  the conformal blocks are even under $z\leftrightarrow\bar z$ exchange, that leaves 
$u$ and $v$ unchanged.} 
\be\label{eq:Ansatz0}
G^{(p)}_e(z,\bar z)=\Big(\frac{z \bar z}{z-\bar z}\Big)^{2\,p+1}\,g_e^{(p)}(z,\bar z),\ \ \ \ 
g_e^{(p)}(z,\bar z)\equiv \sum_{m,n}c_{m,n}^e\mathcal{F}_{\rho_1+m,\,\rho_2+n}^{-\;(a_e,b_e;c_e)}(z,\bar z),
\ee
where $c_{m,n}^e$ are coefficients to be determined and the sum over the two integers $m$ and $n$ in eq.(\ref{eq:Ansatz0}) is so far unspecified. 
Notice that all the functions $\mathcal{F}$ entering the sum over $m$ and $n$ have the same values of $a_e$, $b_e$ and $c_e$.
Matching eq.(\ref{eq:Ansatz0}) in the limit  $z,\bar z \rightarrow 0$ with eq.(\ref{zzbar0limitAnypl}) allows us to determine $\rho_1$ and $\rho_2$, modulo a shift by an integer.
We take
\be 
\rho_1 = \bar \lambda^{(p)}\,,  \ \ \ \ \ \rho_2 = \lambda^{(p)} - p - 1\,,
\ee
in which case the sum over $n$ is bounded from below by $n_{min} = -p$. At this value of $n$, we have $m(n_{min}) = e-p$. 
There is no need to discuss separately the behaviour of the blocks with $\ell\leq p$. Their form is still included in the ansatz (\ref{eq:Ansatz0}) with the additional
requirement that some coefficients $c^e_{m,n_{min}}$ should vanish. This condition is automatically satisfied in the final solution. 
In the next subsections we will discuss the precise range of the sum over $m$ and $n$ and explain how the coefficients $c_{m,n}^e$ can be determined.


\subsection{Reduction to a Linear System}

\label{subsec:reduction}

The eigenfunctions $\mathcal{F}_{\rho_1,\,\rho_2}^{\pm\;(a,b;c)}(z,\bar z)$ have several properties that would allow us to find a solution to the system (\ref{eq:CasimirSystem}). 
In order to exploit such properties, we first have to express the system (\ref{eq:CasimirSystem}) for $G_e^{(p)}$  in terms of the functions $g_e^{(p)}(z,\bar z)$ defined in eq.(\ref{eq:Ansatz0}).
We plug the ansatz (\ref{eq:Ansatz0}) in eq.(\ref{eq:CasimirSystem}) and use the following relations
\bea
\Delta^{(a,b;c)}_\epsilon\Big(\frac{z \bar z}{z-\bar z}\Big)^{k} & = & \Big(\frac{z \bar z}{z-\bar z}\Big)^{k} \bigg( \Delta^{(a,b;c)}_{\epsilon-2k}+k\,(k-\epsilon+c-1) - k\,(k-\epsilon+1)\, \frac{z\bar z(z+\bar z) -2z\bar z }{(z-\bar z)^2} \bigg)\,, \nn \\
L(\mu)\Big(\frac{z \bar z}{z-\bar z}\Big)^{k}& = &\Big(\frac{z \bar z}{z-\bar z}\Big)^{k}
\bigg(L(\mu) + k\;\frac{z+\bar z-2 z\bar z}{(z-\bar z)^2}\bigg)\,,
\label{DeltaLrelations}
\eea
to obtain the system of Casimir equations for $g_e^{(p)}$:
\be\label{eq:SystemCasimir_modified} 
\widetilde{Cas}_e^{(p)}(g) \equiv Cas^0\, g_{e}^{(p)} + Cas^{+}\, g_{e+1}^{(p)} +Cas^{-}\, g_{e-1}^{(p)}=0\,.
\ee
We have split each Casimir equation in terms of three differential operators $Cas^0$, $Cas^+$, $Cas^-$, that act
on $g_e^{(p)}$,  $g_{e+1}^{(p)}$ and  $g_{e-1}^{(p)}$, respectively. In order to avoid cluttering, we have omitted the obvious 
$e$ and $p$ dependences of such operators. Their explicit form is as follows:
\bea
Cas^0 & = & 
\Big(\frac{z-\bar z}{z\bar z}\Big)^2 \Big(\Delta_{0}^{(a_e,b_e;c_e)}+(1+2p)(2p-2-e)-\frac{1}{2}\,\big(E_{\ell}^{p}-\varepsilon_{e}^{p}\big)\Big) \nn \\
&& -3p\,\frac{z-\bar z}{z\bar z}\times \Big((1-z)\partial_z - (1-\bar z)\partial_{\bar z}\Big)
 -p\,(1+2p)\,\frac{z+\bar z-2}{z\bar z},
\label{eq:casimir_modifed_part0} \\
Cas^{+}& = & 
B_{e}\,\frac{z-\bar z}{z\bar z}\times\frac{z-\bar z}{z\bar z}L(b_{e+1})
+(1+2p)\,B_{e}\,\frac{z+\bar z-2z\bar z}{z\bar z} \frac{1}{z\bar z},
 \label{eq:casimir_modifed_part+} \\
Cas^{-}&=&
A_{e}^p\, \frac{z-\bar z}{z\bar z}\times(z-\bar z) L(a_{e-1})
+(1+2p)\,A_{e}^p\,\frac{z+\bar z-2z\bar z}{z\bar z}.\label{eq:casimir_modifed_part-}
\eea
Notice that the action of $\Delta_{0}^{(a_e,b_e;c_e)}$ in eq.(\ref{eq:casimir_modifed_part0}) on $g_{e}^{(p)}$ is trivial and gives just the sum of the eigenvalues of the $\mathcal{F}_{\rho_1,\,\rho_2}^{-\;(a,b;c)}(z,\bar z)$ entering $g_{e}^{(p)}$.
It is clear from the form of the ansatz (\ref{eq:Ansatz0}) that the system (\ref{eq:SystemCasimir_modified}) involves three different kinds of
functions $\mathcal F^-$, with different values of $a$, $b$ and $c$ (actually only $b$ and $c$ differ, recall eq.(\ref{abcExp})).

\begin{figure}[!t]
\vspace{0.5cm}
\begin{center}
\begin{minipage}{0.3\linewidth}
\begin{center}
        \fbox{\footnotesize ${\cal R}_0$} \\[0.05cm]
        \includegraphics[width=40mm]{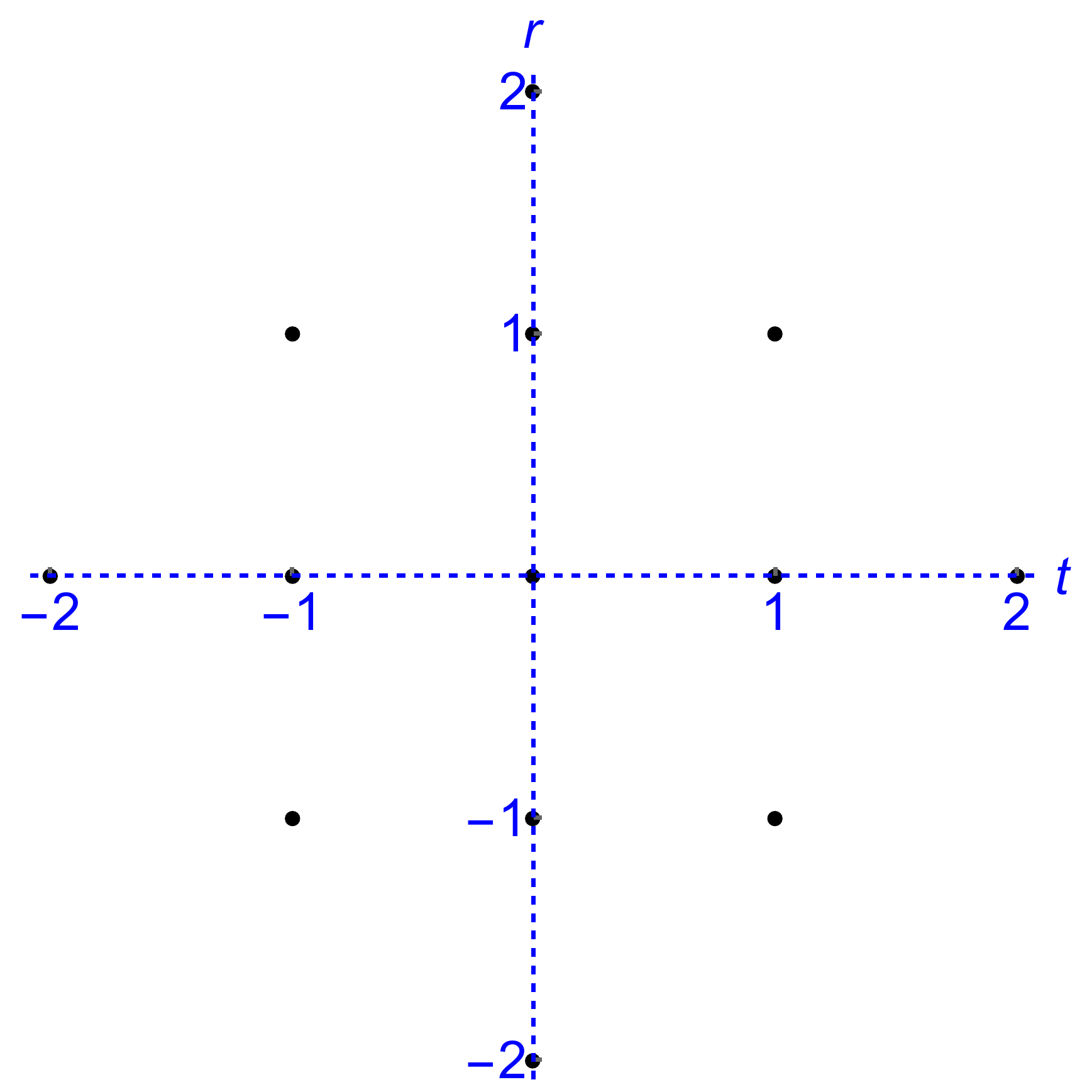}\\
\end{center}
\end{minipage}
\hspace*{0.3cm}
\begin{minipage}{0.3\linewidth}
\begin{center}
        \fbox{\footnotesize ${\cal R}_+$} \\[0.05cm]
        \includegraphics[width=40mm]{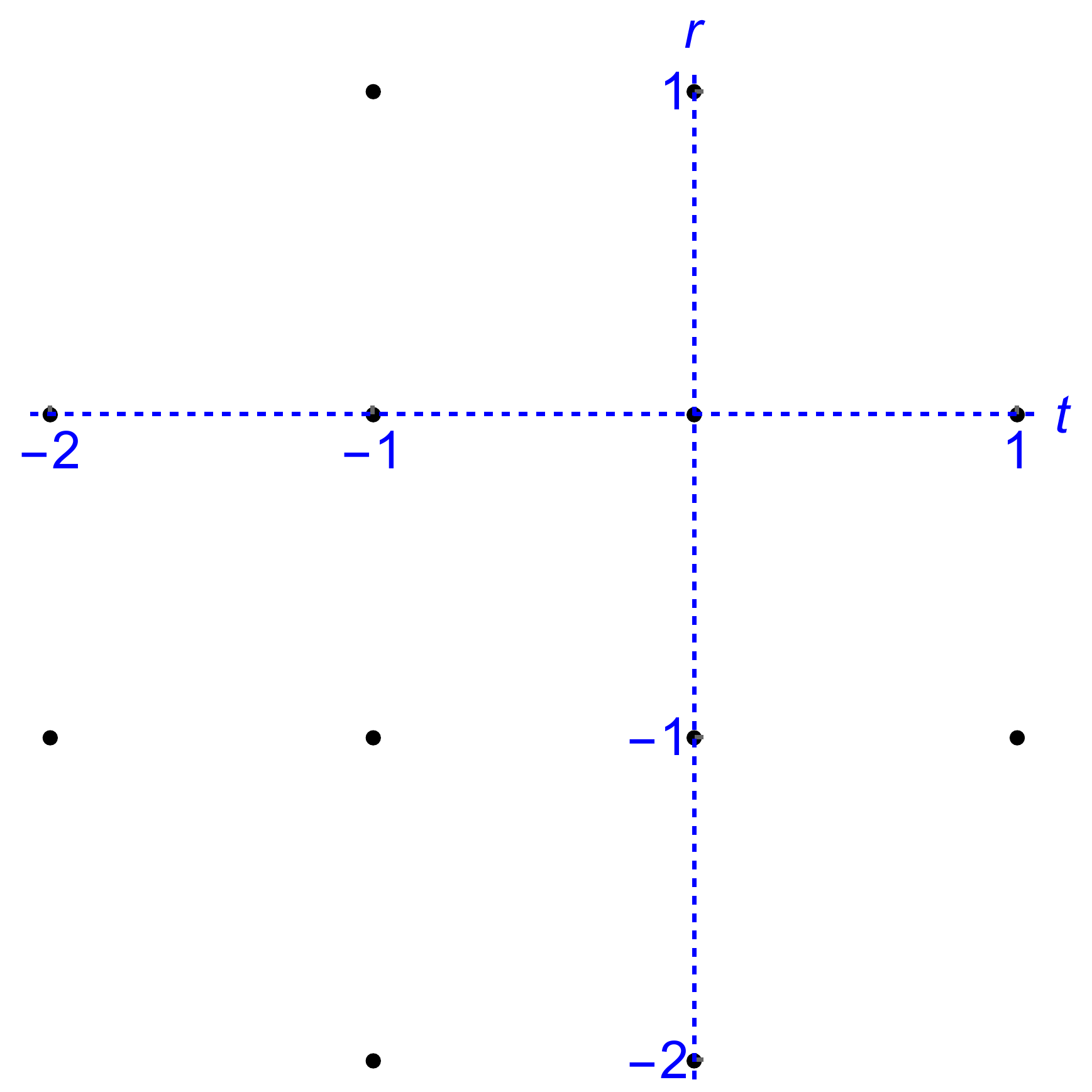}\\
\end{center}
\end{minipage}
\hspace*{0.3cm}
\begin{minipage}{0.3\linewidth}
\begin{center}
        \fbox{\footnotesize ${\cal R}_-$} \\[0.05cm]
        \includegraphics[width=40mm]{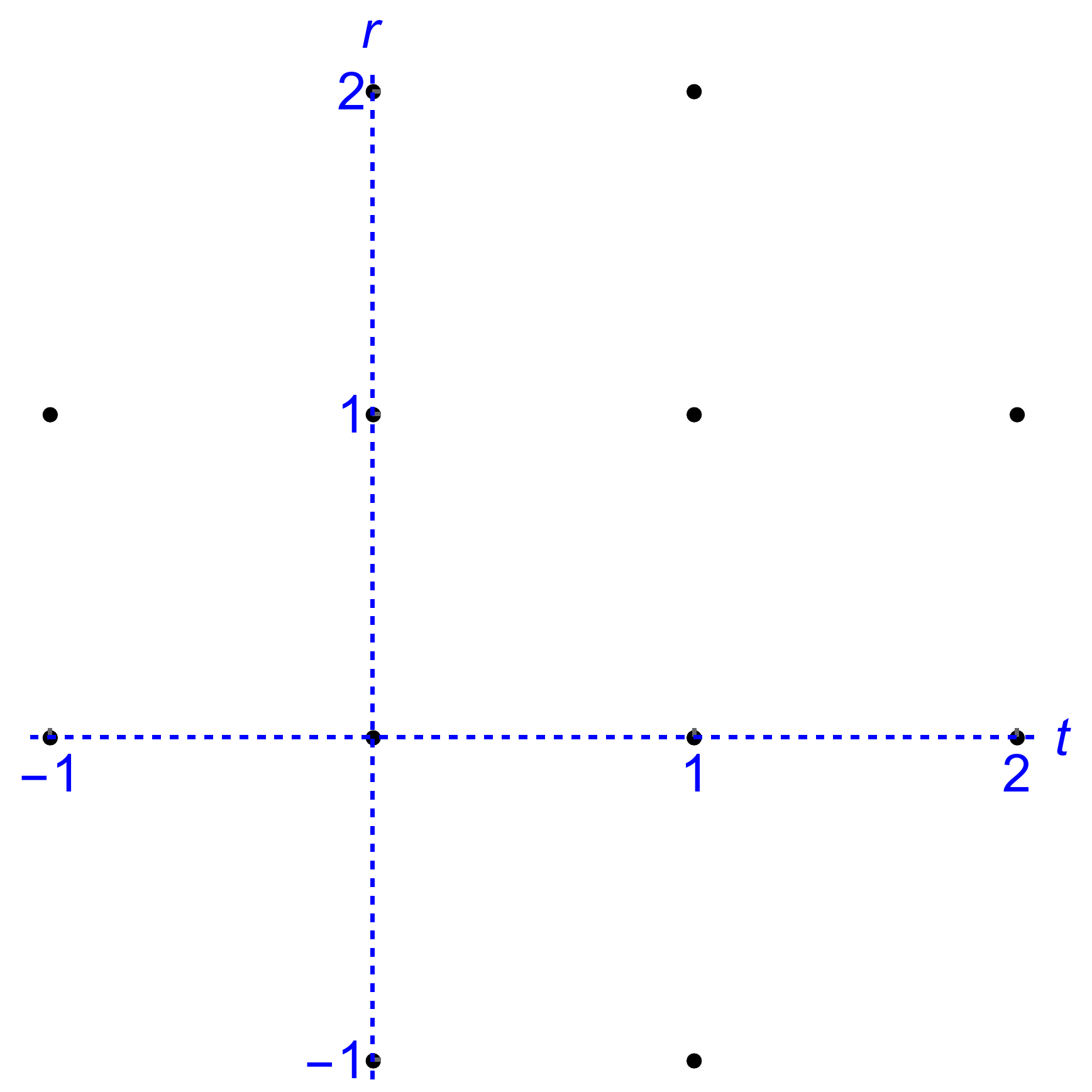}\\
\end{center}
\end{minipage}
\end{center}
\vspace*{-0.2cm}
\caption{\label{fig:Rpm0}
Set of points in the $(r,t)$ plane forming the regions ${\cal R}_0$ (13 points),  ${\cal R}_+$ (12 points) and  ${\cal R}_-$ (12 points) defined in eqs.(\ref{eq:properties_group1})-(\ref{eq:properties_group3}).}
\end{figure}

Using properties of hypergeometric functions, however, we can bring the Casimir system (\ref{eq:SystemCasimir_modified}) into an algebraic system 
involving functions $\mathcal{F}_{\rho_1+r,\,\rho_2+t}^{-\;(a_{e},b_{e};c_{e})}(z,\bar z)$ only, with different values of $r$ and $t$, but crucially with the same values
of $a_e$, $b_e$ and $c_e$. In order to do that, it is useful to interpret  
each of the terms entering  the definitions of $Cas^0$, $Cas^+$ and $Cas^-$ as an operator acting on the functions $\mathcal F^-$ shifting their parameters.
Their action can be reconstructed from the more fundamental operators provided in the appendix \ref{app:properties}.
For each function ${\mathcal F}^-$ appearing in the ansatz (\ref{eq:Ansatz0}), we have
\bea
\label{eq:properties_group1}
Cas^0 \,\mathcal{F}_{\rho_1+m,\,\rho_2+n}^{\;-(a,b;c)}(z,\bar z)& = & \sum_{(r,t)\in {\cal R}_0}\,A^0_{r,t}(m,n)\,\mathcal{F}_{\rho_1+m+r,\,\rho_2+n+t}^{-\;(a,b;c)}(z,\bar z)\,, \\
\label{eq:properties_group2} 
Cas^+\,\mathcal{F}_{\rho_1+m,\,\rho_2+n}^{-\;(a,b;c)}(z,\bar z)& = & \sum_{(r,t)\in {\cal R}_+}\,A^+_{r,t}(m,n)\,\mathcal{F}_{\rho_1+m+r,\,\rho_2+n+t}^{-\;(a,b+1;c+1)}(z,\bar z)\,,  \\
\label{eq:properties_group3}
Cas^-\,\mathcal{F}_{\rho_1+m,\,\rho_2+n}^{-\;(a,b;c)}(z,\bar z)& =&\sum_{(r,t)\in {\cal R}_-}\,A^-_{r,t}(m,n)\,\mathcal{F}_{\rho_1+m+r,\,\rho_2+n+t}^{-\;(a,b-1;c-1)}(z,\bar z)\,,
\eea
where $A^0$, $A^-$ and $A^+$ are coefficients that in general depend on all the parameters involved: $a$, $b$, $\Delta$, $\ell$, $e$ and $p$ but not on $z$ and $\bar z$, namely they are just constants.
For future purposes, in eqs.(\ref{eq:properties_group1})-(\ref{eq:properties_group3}) we have only made explicit the dependence  of $A^0$, $A^-$ and $A^+$ on the integers $m$ and $n$.
The sum over $(r,t)$ in each of the above terms runs over a given set of pairs of integers. We report in fig.~\ref{fig:Rpm0} the values of $(r,t)$ spanned in each of the three regions ${\cal R}_0$, ${\cal R}_+$ and ${\cal R}_-$.
We do not report the explicit and quite lengthy expression of the coefficients $A^0_{r,t}$, $A^+_{r,t}$ and $A^-_{r,t}$, but we refer the reader again to appendix \ref{app:properties} where we provide all the necessary
relations needed to derive them.
Using eqs.(\ref{eq:Ansatz0}) and (\ref{eq:properties_group1})-(\ref{eq:properties_group3}),  the Casimir system (\ref{eq:SystemCasimir_modified}) can be rewritten in terms of the functions 
 $\mathcal{F}^-$ only, with the same set of coefficients $a_{e}$, $b_{e}$ and $c_{e}$:\footnote{It is understood that $c_{m,n}^{-1} = c_{m,n}^{p+1}=0$ in eq.(\ref{fullsystem}).}
 \be
 \sum_{m,n} \Big( \hspace{-0.15cm}\sum_{(r,t)\in {\cal R}_0} \hspace{-0.25cm}A^0_{r,t}(m,n)\, c_{m,n}^e 
+  \hspace{-0.15cm}\sum_{(r,t)\in {\cal R}_+}\hspace{-0.25cm} A^+_{r,t}(m,n) \,c_{m,n}^{e+1} 
+  \hspace{-0.15cm}\sum_{(r,t)\in {\cal R}_-} \hspace{-0.25cm}A^-_{r,t}(m,n)\, c_{m,n}^{e-1} \Big)
\mathcal{F}_{\rho_1+m+r,\,\rho_2+n+t}^{-\;(a_e,b_e;c_e)} =0
  \label{fullsystem}  \,.
\ee
The functions ${\mathcal F}^-$ appearing in eq.(\ref{fullsystem}) are linearly independent among each other, since they all have a different asymptotic behaviour as  $z,\bar z\rightarrow 0$. 
Hence the only way to satisfy eq.(\ref{fullsystem}) is to demand that terms multiplying different ${\mathcal F}^-$  vanish on their own:
\bea
 && \sum_{(r,t)\in {\cal R}_0} \! A^0_{r,t}(m^\prime-r,n^\prime-t) c_{m^\prime-r,n^\prime-t}^e 
+  \sum_{(r,t)\in {\cal R}_+} \!A^+_{r,t}(m^\prime-r,n^\prime-t) c_{m^\prime-r,n^\prime-t}^{e+1} \nn \\
&&+ \sum_{(r,t)\in {\cal R}_-} \!A^-_{r,t}(m^\prime-r,n^\prime-t) c_{m^\prime-r,n^\prime-t}^{e-1}=0\,, \ \ \ \ \forall m^\prime,n^\prime, \ \ e=0,\ldots p
  \label{eq:inear_algebraic_system}  \,,
\eea
where $m^\prime=m+r$, $n^\prime=n+t$.
The Casimir system is then reduced to the over-determined {\it linear algebraic system of equations} (\ref{eq:inear_algebraic_system}).


\subsection{Solution of the System}

\begin{figure}[!t]
\vspace{0.5cm}
\begin{center}
\hspace{0.25cm}
\begin{minipage}{0.5\linewidth}
\begin{center}
        \fbox{\footnotesize $e=0$} \\[0.05cm]
        \includegraphics[width=60mm]{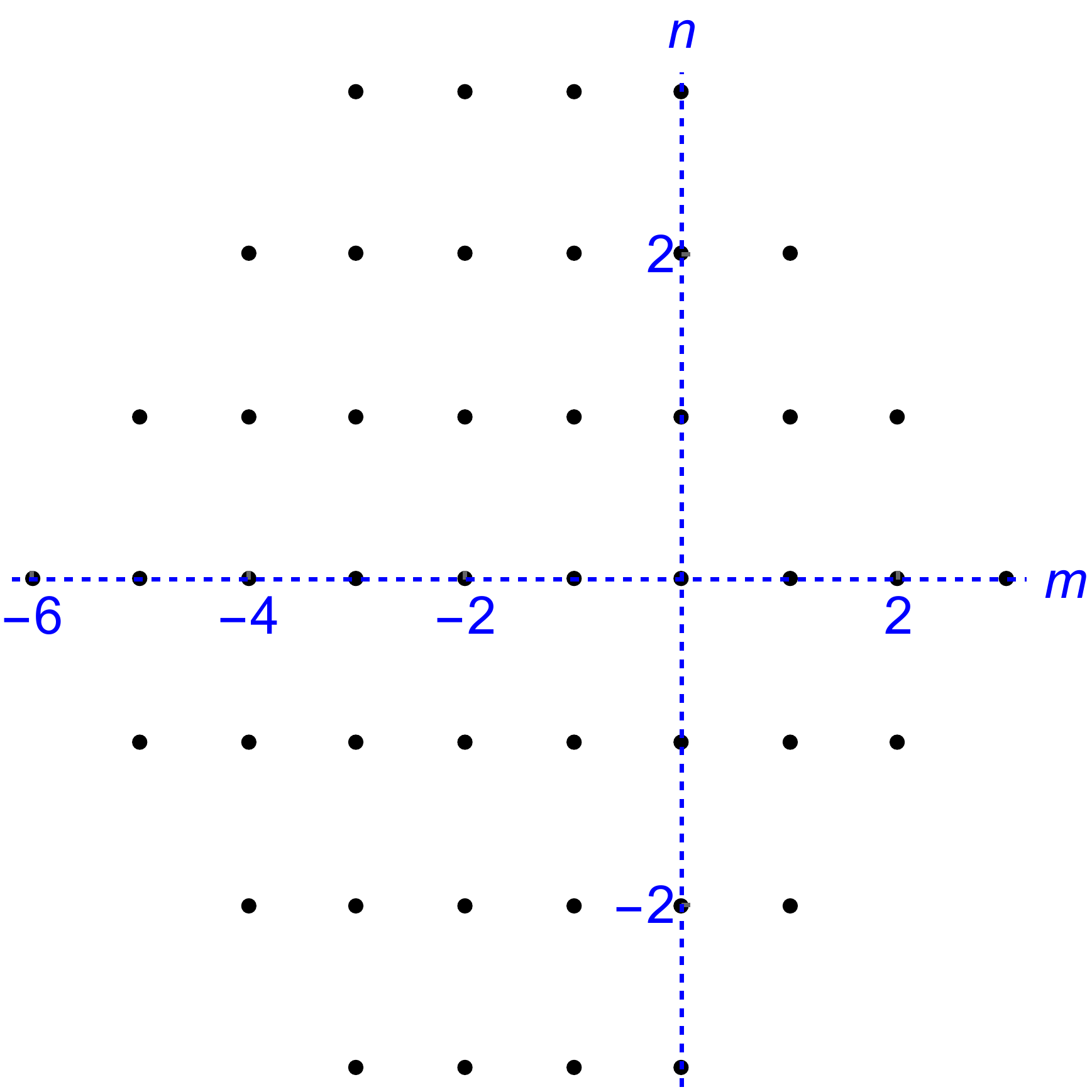}\\
\end{center}
\end{minipage}
\hspace*{-0.6cm}
\begin{minipage}{0.5\linewidth}
\begin{center}
        \fbox{\footnotesize $e=1$} \\[0.05cm]
        \includegraphics[width=60mm]{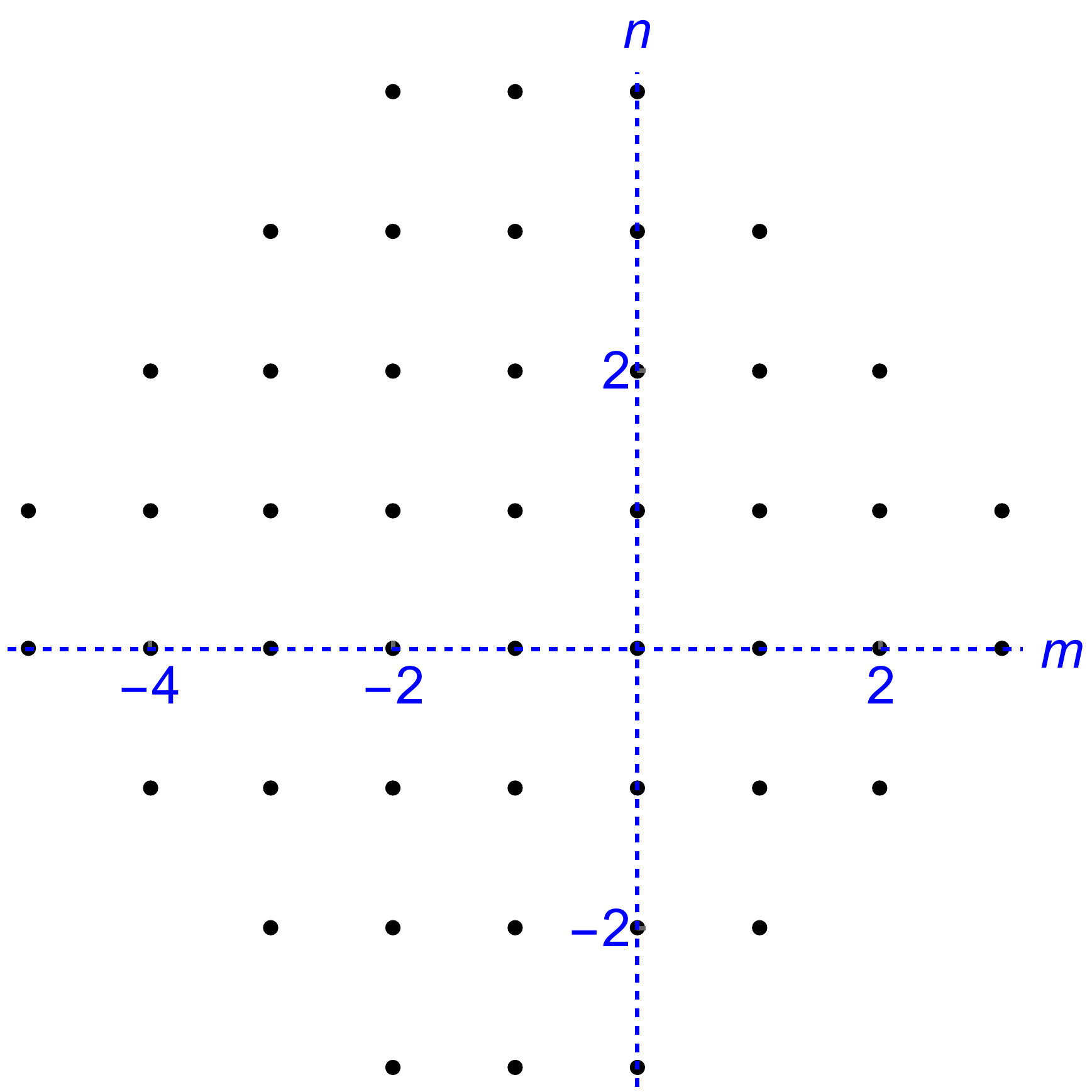}\\
\end{center}
\end{minipage}
\\[0.65cm]
\hspace{0.25cm}
\begin{minipage}{0.5\linewidth}
\begin{center}
        \fbox{\footnotesize $e=2$} \\[0.05cm]
        \includegraphics[width=60mm]{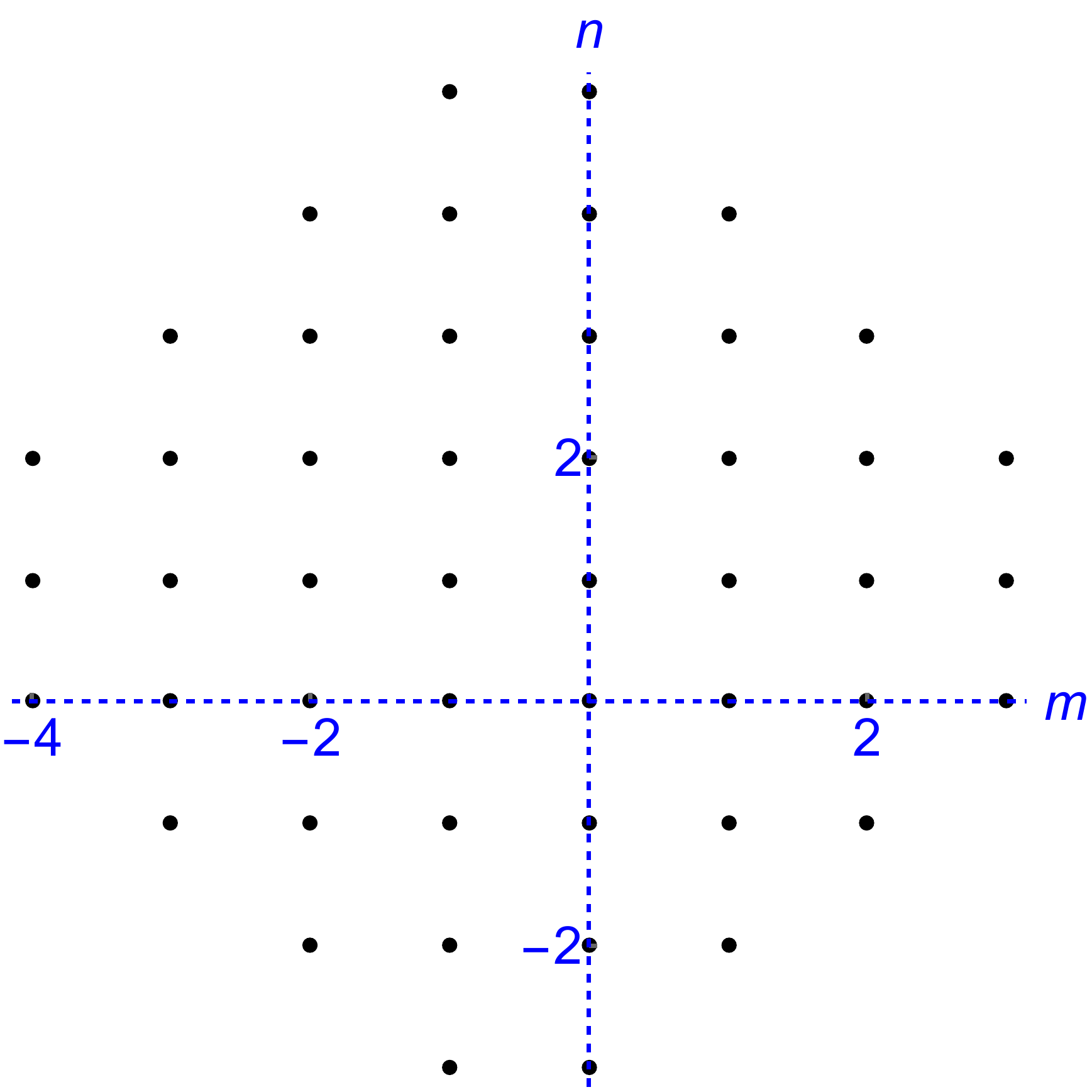}\\
\end{center}
\end{minipage}
\hspace*{-0.6cm}
\begin{minipage}{0.5\linewidth}
\begin{center}
        \fbox{\footnotesize $e=3$} \\[0.05cm]
        \includegraphics[width=60mm]{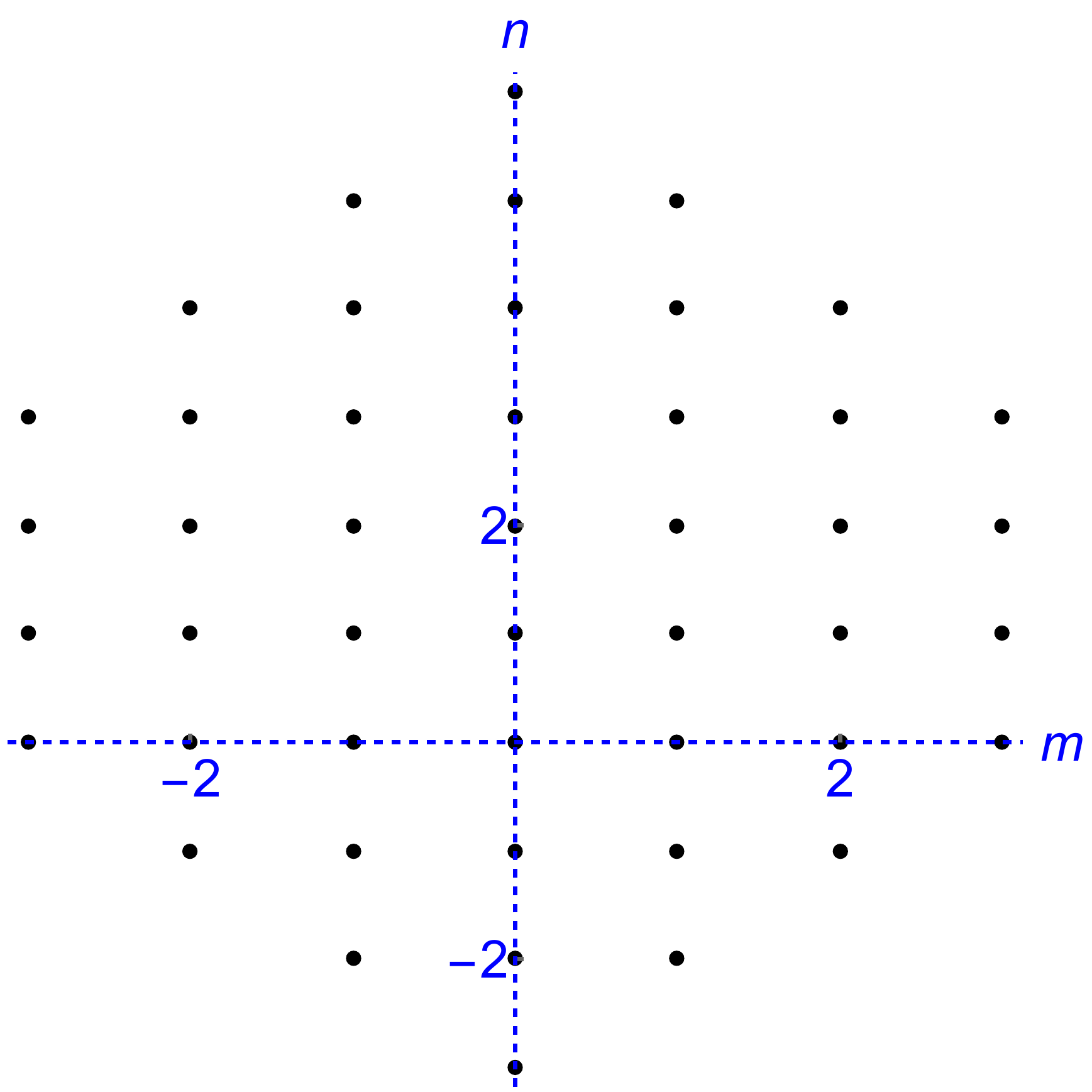}\\
\end{center}
\end{minipage}
\end{center}
\vspace*{-0.2cm}
\caption{\label{fig:Octagons}
Set of non-vanishing coefficients $c_{m,n}^e$ (represented as black dots) entering the ansatz for mixed tensor CBs  in eq.(\ref{eq:Ansatz}) for $p=3$ and $e=0,1,2,3$. For $e=0$ and $e=p$
the octagons collapse to  hexagons.}
\end{figure}

In order to solve the system (\ref{eq:inear_algebraic_system}), we have to determine the range of values of $(m,n)$ entering the ansatz
(\ref{eq:Ansatz0}), that also determines the size of the linear system. 
By rewriting the known $p=1$ and $p=2$ CBs found using the shadow formalism in the form of eq.(\ref{eq:Ansatz0}), 
we have deduced the range in $(m,n)$ of the coefficients $c_{m,n}^e$
for any $p$ (a posteriori proved using the results below).
\begin{figure}[!t]
    \centering
    \includegraphics[width=0.8\textwidth]{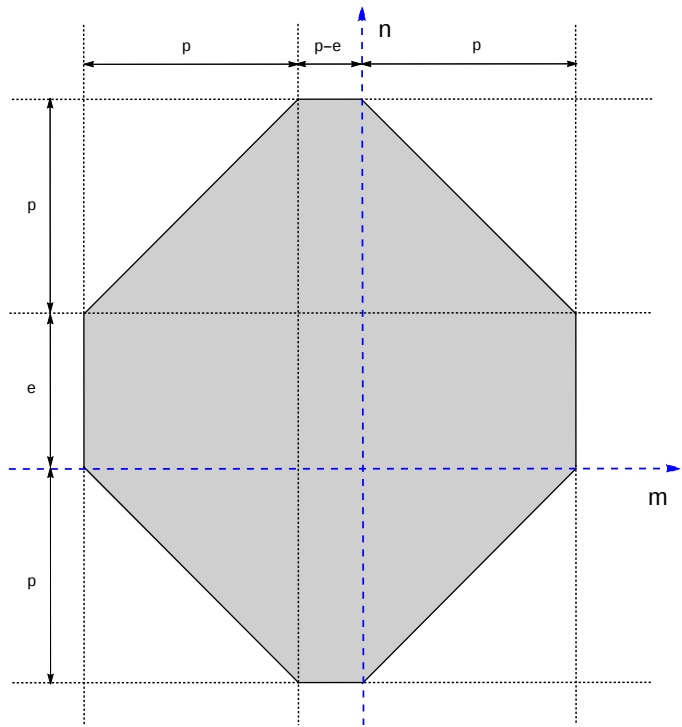}
     \caption{The dimensions of the generic octagon enclosing the lattice of non-vanishing coefficients $c_{m,n}^e$ entering the ansatz for mixed tensor CBs  in eq.(\ref{eq:Ansatz}).}
    \label{fig:octagon_dimensions}
\end{figure}
For each value of $e$, the non-trivial coefficients $c_{m,n}^e$ span a two-dimensional lattice in the $(m,n)$ plane. For each $e$, the shape of the lattice is an octagon, with $p$ and $e$ dependent edges.
The position and shape of the generic octagon in the $(m,n)$ plane is depicted in fig.~\ref{fig:octagon_dimensions}. One has
\begin{equation}\label{octagon}
n_{min}=-\,p,\;\;n_{max}=e+p,\;\;m_{min}=e-2\,p,\;\;m_{max}=p\,.
\end{equation}
For $e=0$ and $e=p$, the octagons collapse to hexagons. 
The number $N_p^e$ of points inside a generic octagon is
\begin{equation}\label{NepDef}
N_p^e=2p\,(2p-e)+(1+e)\,(3p+1-e)
\end{equation}
and correspond to the number of non-trivial coefficients $c_{m,n}^e$ entering the ansatz (\ref{eq:Ansatz0}).
The total number $N_p$ of coefficients to be determined at level $p$ is then
\be\label{eq:number_coefficients}
N_p\equiv\sum_{e=0}^p\,N_p^e=(1+p)\,\Big(1+\frac{17}{6}\,p+\frac{25}{6}\,p^2\Big)\,.
\ee
The size of the linear system grows as $p^3$.
The first values are $N_1= 16$, $N_2=70$, $N_3=188$, $N_4=395$.
For illustration, we report in fig.~\ref{fig:Octagons} the explicit lattice of non-trivial coefficients $c_{m,n}^e$ for $p=3$.

The system (\ref{eq:inear_algebraic_system}) is always over-determined, since it is spanned by the values $(m',n')$ whose range is bigger than the range of $(m,n)\in Oct_e^{(p)}$ (spanning all the coefficients to be determined) due to the presence of $(r,t)\in [-2,2].$ There are only $N_p-1$ linearly independent equations,  because the system of Casimir equations can only determine conformal blocks up to an overall factor. 
The most important property of the system  (\ref{eq:inear_algebraic_system}) is the following: while the number of equations grows with $p$, the total number of coefficients $c_{m,n}^e$ entering any given equation 
in the system (\ref{eq:inear_algebraic_system}) does not.
This is due to the ``local nearest-neighbour" nature of the interaction between the blocks, for which at most three conformal blocks can enter the Casimir system (\ref{eq:CasimirSystem}), independently of the value of $p$.
More precisely,  all the equations (\ref{eq:inear_algebraic_system})
involve from a minimum of one coefficient $c_{m,n}^e$ up to a maximum of 37 ones. Thirty seven corresponds to the total number of coefficients $A^0$, $A^+$ and $A^-$
entering eqs.(\ref{eq:properties_group1})-(\ref{eq:properties_group3}), see fig.\ref{fig:Rpm0}.
The only coefficients that enter alone in some equations are the ones corresponding to the furthermost vertices of the hexagons, namely 
\be\label{specialC}
c_{0,-p}^p, c_{0,2p}^p, c_{p,0}^0, c_{-2p,0}^0\,.
\ee
For instance, let us take $n^\prime=-2-p$ and $e=p$ in eq.(\ref{eq:inear_algebraic_system}), with $m^\prime$ generic.
Since $n_{min} = -p$, a non-vanishing term can be obtained only by taking $t=-2$. Considering that $c^{p+1}=0$ and ${\cal R}_-$ does not include $t=-2$ (see fig.\ref{fig:Rpm0}),
this equation reduces to 
\be \label{eq:A0Exp}
A^0_{0,-2}(m,-p)|_{e=p} \; c_{m,-p}^p = 0 \,, \ \ \ \forall m\,,
\ee
where $m^\prime=m$, since the point in ${\cal R}_0$ with $t=-2$ has $r=0$.
This equation forces all the coefficients $c_{m,-p}^p$ to vanish, unless the factor $A^0_{0,-2}(m,-p)$ vanishes on its own.
One has
\be
A^0_{0,-2}(m,n)|_{e=p}\propto (m+n+p)\Delta +(m-n-p)\ell + m^2+\frac12 m (p-2)+(n+p)(n+\frac 32p-2)\nn\,.
\ee
This factor is generally non-vanishing, unless $m=0$ and $n=-p$, in which case it vanishes for any $\Delta$, $\ell$ and $p$.
In this way eq.(\ref{eq:A0Exp}) selects $c_{0,-p}^p$ as the only non-vanishing coefficient at level $n=-p$ for $e=p$.
Notice that it is crucial that  $A^0_{0,-2}(m,n)|_{e=p}$ vanishes automatically for a given pair ($m,n$), otherwise either the whole set of equations would only admit the trivial
solution $c_{m,n}^e=0$, or the system would be  infinite dimensional.
A similar reasoning applies for the other three coefficients. One has in particular
\bea
A^0_{0,2}(0,2p)|_{e=p} && c_{0,2p}^p  =  0 \,, \nn \\
A^0_{2,0}(p,0)|_{e=0} && c_{p,0}^0  =  0 \,, \label{eq:A0220} \\
A^0_{-2,0}(-2p,0)|_{e=0} && c_{-2p,0}^0 =  0 \,,  \nn
\eea
that are automatically satisfied because the three coefficients $A^0_{0,2}$, $A^0_{2,0}$ and $A^0_{-2,0}$ vanish when evaluated
for the specific values reported in eq.(\ref{eq:A0220}) for any $\Delta$, $\ell$ and $p$.

The system~(\ref{eq:inear_algebraic_system}) is efficiently solved by extracting a subset of $N_p-1$ linearly independent  equations. 
This can be done by fixing the values $(r,t)=(r^*,t^*)$ entering the definitions of $(m',n')$. There are 4 very special subsets of the $N_p-1$ equations (corresponding to very specific values $(r^*,t^*)$) which allows us to determine the solution iteratively starting from eq.(\ref{eq:inear_algebraic_system}).
They correspond to a solution where one of the four coefficients (\ref{specialC}) is left undetermined, in other words $(r^*,t^*)$ can be set to be $(0,-2)$, $(0,2)$, $(2,0)$ or $(-2,0)$.
For instance, if we choose $c_0\equiv c_{0,-p}^p$ as the undetermined coefficient, a recursion relation is found from eq.(\ref{eq:inear_algebraic_system}) by just singling out the term with $t=-2$
in $A^0$ and setting $(r^*,t^*)=(0,-2)$. Such a choice leads to $m^\prime= m$, $n^\prime= n-2$, and one finally gets
\bea
-A^0_{0,-2}(m,n) c_{m,n}^e &=&  \sum_{\substack{(r,t)\in {\cal R}_0  \\
       (r,t)\neq (0,-2)}} A^0_{r,t}(m-r,n-2-t) c_{m-r,n-2-t}^e \nn \\
&+&  \sum_{(r,t)\in {\cal R}_+} A^+_{r,t}(m-r,n-2-t) c_{m-r,n-2-t}^{e+1} \label{recursivesol} \\
&+& \sum_{(r,t)\in {\cal R}_-} A^-_{r,t}(m-r,n-2-t) c_{m-r,n-2-t}^{e-1} \nn  \,.
\eea
It is understood in eq.(\ref{recursivesol}) that $c_{m,n}^e=0$ if the set $(m,n)$ lies outside the $e$-octagon of coefficients.
The recursion (\ref{recursivesol}) allows us to determine all the coefficients $c_{m,n}^e$ at a given $e=e_0$ and $n=n_0$ in terms of the ones $c^e_{m,n}$ with $n<n_0$ and $c^e_{m,n_0}$ with $e>e_0$.
Hence, starting from $c_0$, one can determine all $c_{m,n}^e$ as a function of $c_0$ for any $p$. The overall normalization of the CBs is clearly irrelevant
and can be reabsorbed in a redefinition of the OPE coefficients. However, some care should be taken in the choice of $c_0$ if one wants to avoid  the appearance of spurious divergencies in the CBs for specific values 
of $\ell$ and $\Delta$. These divergencies are removed by a proper $\Delta$ and $\ell$ dependent rescaling of $c_0$.
From eq.(\ref{eq:inear_algebraic_system}) one can easily write the three other relations similar to eq.(\ref{recursivesol}) to determine recursively $c_{m,n}^e$ starting from 
$c_{0,2p}^p$, $c_{p,0}^0$ or $c_{-2p,0}^0$.

We can finally write down the full analytic solution for the CBs $G_e^{(p)}$:
\be\label{eq:Ansatz}
G^{(p)}_e(z,\bar z)=\Big(\frac{z \bar z}{z-\bar z}\Big)^{2\,p+1}\!\! \!\! \!\! \sum_{(m,n)\in Oct_e^{(p)}} \! \!\! c_{m,n}^e\mathcal{F}_{\frac{\Delta+\ell+\frac p2}2+m,\,\frac{\Delta-\ell+\frac p2}2-(p+1)+n}^{-\;(a_e,b_e;c_e)}(z,\bar z),
\ee
where $c_{m,n}^e$ satisfy the recursion relation (\ref{recursivesol}) (or any other among the four possible ones) and $(m,n)$ runs over the points within the $e$-octagon depicted in fig.\ref{fig:octagon_dimensions}.

A similar analysis can be performed for the conjugate blocks $\overline G_e^{(p)}$. We do not report here the detailed derivation that is logically identical to the one above, but just the final solution:
\be\label{eq:AnsatzConj}
\overline G^{(p)}_e(z,\bar z)=\Big(\frac{z \bar z}{z-\bar z}\Big)^{2\,p+1}\!\! \!\! \!\! \sum_{(m,n)\in Oct_{p-e}^{(p)}} \!\!\! \bar c_{m,n}^e\mathcal{F}_{\frac{\Delta+\ell-\frac p2}2+e+m,\,\frac{\Delta-\ell-\frac p2}2+e-(p+1)+n}^{-\;(a_e,b_e;c_e)}(z,\bar z).
\ee
where
\be
\bar c_{m,n}^e(a,b,\Delta,l,p) =   4^e c_{m,n}^{p-e}\Big(- a +\frac p2,- b -\frac p2,\Delta,l,p\Big)\,.
\ee
Generating the full explicit solution from eq.(\ref{recursivesol}) can be computationally quite demanding for large values of $p$.  
For concreteness, we only report in appendix \ref{app:peq1} the explicit form of the 16 coefficients $c_{m,n}^e$ for $p=1$ and $a=-b=1/2$.
The general form of $c^e_{m,n}$ for $p=1,2,3,4$ and any $a$, $b$, $\Delta$ and $\ell$ can be downloaded from \href{https://sites.google.com/site/dskarateev/downloads}{https://sites.google.com/site/dskarateev/downloads}.
The blocks $G_e^{(p)}$ for $p=1,2$  and $\overline G_e^{(p)}$ for $p=1$ are in complete agreement with those computed using the shadow formalism. 
By choosing specific values for the parameters $a$ and $b$,  we also have determined the coefficients $c^e_{m,n}$ up to $p=8$, i.e. the value of $p$ that is
obtained in the 4-point function of four energy momentum tensors, see eq.(\ref{pDef}).

It is important to remind the reader that the CBs $G_e^{(p)}$ computed here are supposed to be the seed blocks for possibly other 4-point correlation functions, whose CBs
are determined by acting with given operators on $G_e^{(p)}$ \cite{Echeverri:2015rwa}.
The complexity of the form of the blocks $G_e^{(p)}$ at high $p$ is somehow compensated by the fact that the operators one has to act with become simpler and simpler, the higher is $p$.
An example should clarify the point. Let us consider a 4-point function of spin two operators. In this case, one has to determine conformal blocks associated to the exchange of operators
${\cal O}^{(\ell,\ell+p)}$ (and $\overline{{\cal O}}^{(\ell+p,\ell)}$) for $p=0,2,4,6,8$ (and any $\ell$). The conformal blocks associated to the traceless symmetric operators are obtained by applying up to 8 derivative operators 
in several different combinations to the scalar CB $G_0^{(0)}$. Despite the seed block is very simple, the final blocks are given by (many) complicated sum of derivatives of $G_0^{(0)}$.
The $p=8$ CBs, instead, are essentially determined by the very complicated $G_e^{(8)}$ (and $\overline{G}_e^{(8)}$) blocks, but no significant extra complications come from the external operators.
An example of such phenomenon  in a  four fermion correlator is shown (though in a less significative way)  in section 7.1 of ref.\cite{Echeverri:2015rwa}.
For any given 4-point function, after the use of the differential operators introduced in ref. \cite{Echeverri:2015rwa}, there is no need to compute the coefficients $c_{m,n}^e$ for any $a$ and $b$ but only for
the values of interest. This considerably simplifies the expression of $c_{m,n}^e$.

\subsection{Analogy with Scalar Conformal Blocks in Even Dimensions} 
\label{subsec:scalarCB}

It is worth pointing out in more detail some similarities between the CBs $G_e^{(p)}$ for mixed symmetry tensors computed above and the scalar conformal blocks $G_d$ in $d>2$ even space-time dimensions
($G_4=G_0^{(0)}$ in our previous notation). The quadratic Casimir equation for scalar CBs in any number of dimensions is
\be
\Delta_{d-2}^{(a,b;0)} G_d(z,\bar z) = \frac 12 E_{\ell}(d) G_d(z,\bar z)\,,
\label{CasDiffEqp0}
\ee
where
\be
\label{CasEigenAnyd}
E_{\ell}(d)=\Delta\, (\Delta-d)+\ell (\ell+d-2)
\ee
is the quadratic Casimir eigenvalue for traceless symmetric tensors.
The explicit analytical form of scalar blocks in $d=2,4,6$ dimensions has been found in refs.\cite{Dolan:2000ut,Dolan:2003hv}. 
The same authors also found a relation between scalar blocks in any even space-time dimensionality, eq.(5.4) of ref.\cite{Dolan:2003hv} (see also the more elegant eq.(4.36) of ref.\cite{Dolan:2011dv}),
that allows us to iteratively determine $G_d$ for any $d$, starting from $G_2$.
The $d=4$ and $d=6$ solutions found in ref.\cite{Dolan:2003hv}  have the form
\be\label{eq:object}
G_d (z,\bar z)=  \Big(\frac{z \bar z}{z-\bar z}\Big)^{d-3}  g_d(z,\bar z)\,, \ \ \ \ g_d(z,\bar z)= \sum_{m,n} x_{m,n}\mathcal{F}_{\frac{\Delta+\ell}2+m,\,\frac{\Delta-\ell+2-d}2+n}^{-\;(a,b;0)}(z,\bar z),
\ee
where $a$ and $b$ are as in eq.(\ref{abDef}) with $p=0$ and $x_{m,n}$ are coefficients that in general depend on $\Delta,l,a$ and $b$.
In $d=4$ there is only one non-vanishing coefficient centered at $(m,n)=(0,0)$, while in $d=6$ there are five of them. They are  at $(m,n)=(0,-1)$, $(-1,0)$,
$(0,0)$, $(1,0)$ and $(0,1)$. These five points form a slanted square in the $(m,n)$ plane, centered at the origin.
The explicit form of the coefficients $x_{m,n}$ is known, but it will not be needed in what follows.\footnote{En passant, notice that there is a typo in eq.(2.20) of ref.\cite{Dolan:2003hv} where the block $G_6$  is reported. In the denominator appearing in the last row of that equation, one should replace $(\Delta+\ell-4)(\Delta+\ell-6)\rightarrow (\Delta-\ell-4)(\Delta-\ell-6)$.} 
\begin{figure}[!t]
    \centering
    \includegraphics[width=0.8\textwidth]{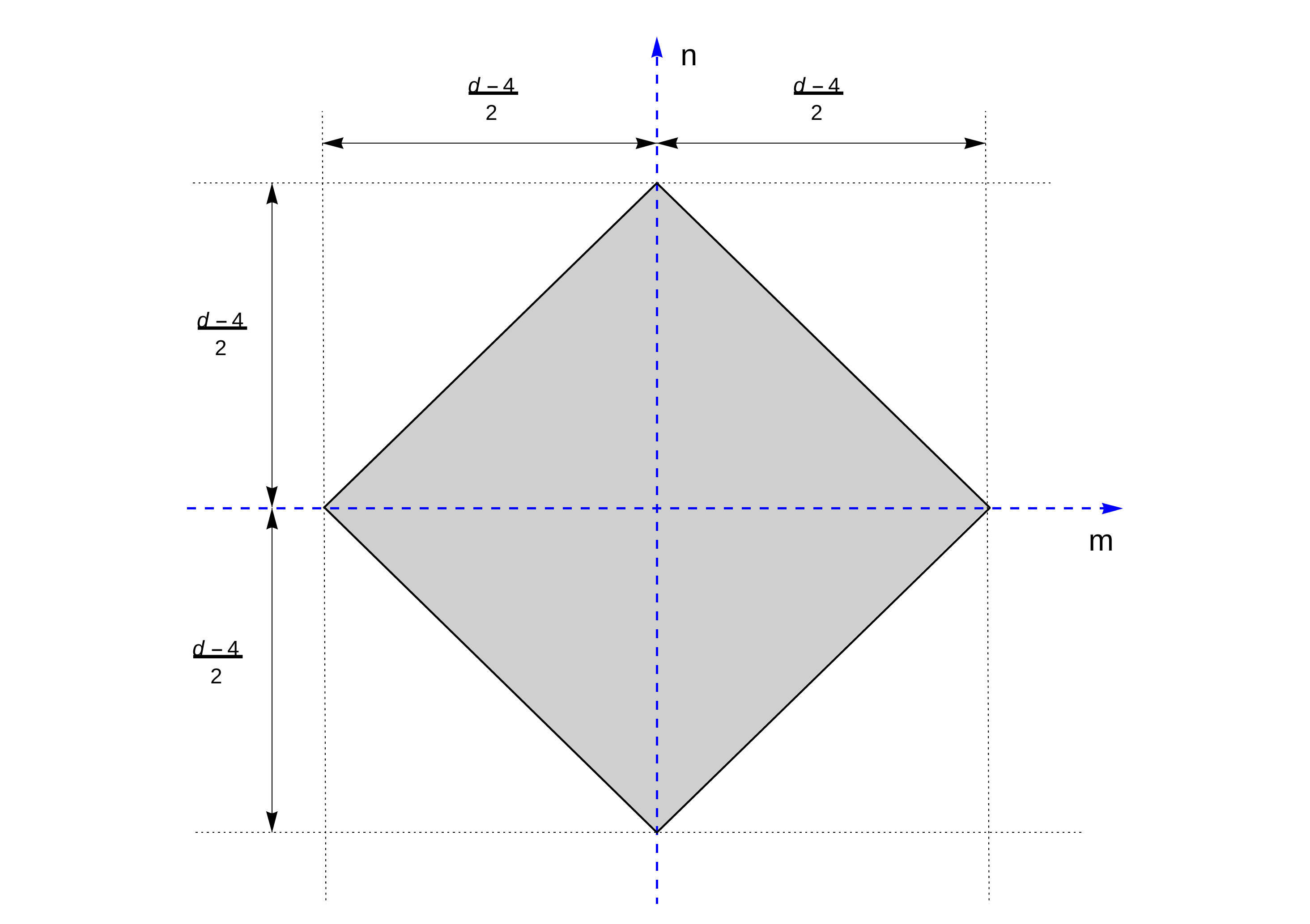}
    \caption{The dimensions of the generic slanted square enclosing the lattice of non-vanishing coefficients $x_{m,n}$ entering the ansatz for scalar symmetric CBs  in eq.(\ref{eq:object}).}
    \label{fig:square_dimensions}
\end{figure}
It is natural to expect that eq.(\ref{eq:object})  should apply for any even $d\geq 4$, with a number of non-vanishing coefficients that increases with $d$.\footnote{See also ref.\cite{Behan:2014dxa}, where similar considerations were conjectured.}
This is not difficult to prove. From the first relation in eq.(\ref{DeltaLrelations}) we can get the form of the Casimir equation for the function $g_d(z,\bar z)$ defined in eq.(\ref{eq:object}), that can be written as
\be
\Big(\frac 1{\bar z}-\frac 1z\Big) \Big(\Delta_{0}^{(a,b;0)} +6 - 2d -\frac 12 E_\ell(d) \Big) g_d = (d-4)  \Big((1-z)\partial_z - (1-\bar z)\partial_{\bar z}\Big) g_d\,.
\label{CasDiffgd}
\ee
Using the techniques explained in subsection \ref{subsec:reduction} and the results of appendix \ref{app:properties}, it is now straightforward to identify which is the range of $(m,n)$ of the non-vanishing coefficients $x_{m,n}$ for any $d$ (see 
fig.\ref{fig:square_dimensions}).\footnote{Alternatively, one might use eq.(4.36) of ref.\cite{Dolan:2011dv} to compute $G_d$ and then recast it in the form (\ref{eq:object}).}  In $d$ dimensions, the minimum and maximum values of $m$ and $n$ are given 
by
\be
n_{min} = \frac{4-d}2\,, \ \ \ n_{max} = \frac{d-4}2\,, \ \ \ 
m_{min} = \frac{4-d}2\,, \ \ \ m_{max} = \frac{d-4}2 \,.
\ee
The number $\widetilde N_d$  of coefficients $x_{m,n}$ entering the ansatz (\ref{eq:object}) for scalar blocks in $d$ even space-time dimensions is easily computed by counting the number of lattice points enclosed in the
slanted square. We have
 \be
\widetilde N_d=\frac{d^2}{2}-3d+5\,.
\ee
For large $d$,  $\widetilde N_d\propto d^2$ and matches the behavior of $N_e^p\propto p^2$  for large $p$ in eq.(\ref{NepDef}).

In light of the above analogy between scalar CBs $G_d$ in even $d$ dimensions and mixed tensor CBs $G_e^{(p)}$ in four dimensions, it would be interesting to investigate whether there exist a set of differential operators that links the blocks
$G_e^{(p+1)}$ (or $G_e^{(p+2)}$) to the blocks $G_e^{(p)}$, in analogy to the operator (4.35) of ref.\cite{Dolan:2011dv} relating $G_{d+2}$ to $G_{d}$.
It would be very useful to find, in this or some other way, a more compact expression for the blocks $G_e^{(p)}$.

Let us finally emphasize a technical, but relevant, point where the analogy between $G_d$ in $d$ dimensions and $G_e^{(p)}$ in 4 dimensions does not hold. 
A careful reader might have noticed that in the Casimir equation for $g_d$ the term proportional to $(z+\bar z)-2$, namely the third term in the r.h.s. of the first equation in eq.(\ref{DeltaLrelations}), automatically vanishes. 
Indeed, if we did not know the power $d-3$ in the ansatz (\ref{eq:object}), we could have guessed it by demanding that term to vanish.
On the contrary, no such simple guess seems to be possible for the power $2p+1$ entering $G_e^{(p)}$, given also the appearance of the operator $L$ defined in eq.(\ref{DeltaLrelations}).
As discussed, we have fixed the power $2p+1$ by means of the shadow formalism.

\section{Summary and Conclusions}
\label{sec:conclu}

We have computed in this paper the seed CBs $G_e^{(p)}$ (and $\overline G_e^{(p)}$) associated to the exchange of mixed symmetry  bosonic and fermionic primary operators ${\cal O}^{(\ell,\ell+p)}$ (and $\overline{\cal O}^{(\ell+p,\ell)}$) in the four point functions (\ref{our seed corr}).
We have found a totally general expression for $G_e^{(p)}$ for any $e$, $p$, $\Delta$, $\ell$ and external scaling dimensions, by solving the Casimir set of differential equations, that can be written in the compact form
(\ref{eq:CasimirSystem}). The shadow formalism has been of fundamental assistance to deduce it and also as a useful cross check for the validity of the results.
The final expression for the CBs is given in eq.(\ref{eq:Ansatz}), the most important formula in the paper.
The CBs are expressed in terms of coefficients $c_{m,n}^e$,  that can be determined recursively, e.g. by means of eq.(\ref{recursivesol}).
For each CB, the coefficients $c_{m,n}^e$ span a 2D octagon-shape lattice in the $(m,n)$ plane, with sizes that depend on $p$ and $e$ and increase as $p$ increases.
 We have reported in Appendix \ref{app:peq1} the explicit form of $c_{m,n}^e$ for the simplest case $p=1$. We have not reported the  $c_{m,n}^e$ for higher values of $p$, since their number and complexity grows with $p$. 
Their explicit form up to $p=4$ can be downloaded from a website.
The CBs up to $p=4$ are enough to bootstrap many tensor correlators, including four conserved spin 1 currents. 

Aside from the obvious application in the numerical bootstrap, the knowledge of the CBs $G_e^{(p)}$ should be useful in other contexts.
Having analytical closed expressions for the blocks should be very useful to generalize the so called analytic bootstrap approach \cite{Fitzpatrick:2012yx,Komargodski:2012ek} to tensor correlators.
It would also be interesting to explore holographic interpretations in AdS$_5$ of the CBs $G_e^{(p)}$ and their possible utility 
in the formulation of higher spin theory in AdS$_5$.
It should be interesting, for bootstrap applications, to systematically study the correlation functions associated to the free theories of $(p,0)$ spinor/tensor fields, that admit a particle interpretation \cite{Siegel:1988gd}.
Indeed, aside from the interest per se, this study might give a useful analytically known benckmark point for future bootstrap analysis involving the CBs $G_e^{(p)}$ found in this paper, 
very much like the role that free scalar theories play in actual analytical and
numerical studies.

The somewhat surprisingly simple form of the Casimir system (\ref{eq:CasimirSystem}), where at most three blocks at a time can enter in a sort of local interaction, and
the geometric interpretation of the coefficients $c_{m,n}^e$ in terms of octagons, are perhaps an indication of a more fundamental symmetry principle.
This should hopefully allows us to gain a better understanding of 4D CFTs or at least, less ambitiously,  more compact expressions for the CBs $G_e^{(p)}$.
Even in absence of a would be underlying symmetry, it is well possible that there is a better way to parametrize the blocks  that we might have overlooked.

\section*{Acknowledgments}

A.C.E. thanks Emilio Trevisani and Mauro Valli for useful discussions. 
The work of M.S. was supported by the ERC Advanced Grant no. 267985 DaMESyFla. A.C.E. and M.S. gratefully acknowledge
support from the Weizmann Institute of Science, and the Simons Center for Geometry and Physics, Stony Brook University at which some of the research for this paper was performed.

\appendix

\section{Properties of the ${\mathcal F}$ Functions}

\label{app:properties}

In this Appendix we provide all the properties of the functions $\mathcal{F}_{\rho_1,\,\rho_2}^{\;(a,b;c)}$ needed for the system of Casimir equations and more specifically to derive
eqs.(\ref{eq:properties_group1})-(\ref{eq:properties_group3}). We will not consider the functions $\mathcal{F}_{\rho_1,\,\rho_2}^{\pm\;(a,b;c)}$ here, since their properties can trivially be deduced from the ones below by demanding both sides to be symmetric/anti-symmetric under the exchange $z\leftrightarrow\bar z$. 

The fundamental identities to be considered can be divided in two sets, depending on whether the values $(a,b,c)$ of the functions ${\mathcal F}$ are left invariant or not. The former identities read
\bea
 \Big(\frac{1}z-\frac 12\Big) \mathcal{F}_{\rho_1,\rho_2}^{(a,b;c)} &  = &  \mathcal{F}_{\rho_1-1,\rho_2}^{(a,b;c)} -D_{\rho_1}^{(a,b,c)} \mathcal{F}_{\rho_1,\rho_2}^{(a,b;c)}
+ B_{\rho_1}^{(a,b,c)} \mathcal{F}_{\rho_1+1,\rho_2}^{(a,b;c)} \label{AppI1} \\
\Big(\frac{1}{\bar z}-\frac 12\Big) \mathcal{F}_{\rho_1,\rho_2}^{(a,b;c)} & = &  \mathcal{F}_{\rho_1,\rho_2-1}^{(a,b;c)} -D_{\rho_2}^{(a,b,c)} \mathcal{F}_{\rho_1,\rho_2}^{(a,b;c)}
+ B_{\rho_2}^{(a,b,c)} \mathcal{F}_{\rho_1,\rho_2+1}^{(a,b;c)}  \label{AppI2}\\
L_0   \mathcal{F}_{\rho_1,\rho_2}^{(a,b;c)} & = &  \rho_2 \mathcal{F}_{\rho_1,\rho_2-1}^{(a,b;c)}- \rho_1 \mathcal{F}_{\rho_1-1,\rho_2}^{(a,b;c)}-(\rho_2+c-1) B_{\rho_2}^{(a,b,c)}\mathcal{F}_{\rho_1,\rho_2+1}^{(a,b;c)}+
  \label{AppI3} \\
&& (\rho_1+c-1) B_{\rho_1}^{(a,b,c)}\mathcal{F}_{\rho_1+1,\rho_2}^{(a,b;c)}+   \frac{1}{2}\,(2-c)(D_{\rho_1}^{(a,b,c)}- D_{\rho_2}^{(a,b,c)} )\mathcal{F}_{\rho_1,\rho_2}^{(a,b;c)} ,  \nn
\eea
where $L_0= \Big((1-\bar z)\partial_{\bar z}-(1-z)\partial_z \Big)$ and we have defined 
\bea
C_{\rho}^{(a,b,c)} & = & \frac{(a+\rho)(b-c-\rho)}{(c+2\rho)(c+2\rho-1)}  \,,  \label{ABCoeff} \\
B_\rho^{(a,b,c)}  & = & C_{\rho}^{(a,b,c)}\,C_{\rho+1}^{(b-1,a,c-1)} = \frac{(\rho+a)(\rho+b)(\rho+c-b)(\rho+c-a)}{(2\rho+c)^2(c+2\rho+1)(c+2\rho-1)}\,, \nn \\
D_{\rho}^{(a,b,c)} & = & \frac{(2a-c)(2b-c)}{2(c+2\rho)(c+2\rho-2)} \,. 
\eea
The latter identities read
\bea
\mathcal{F}_{\rho_1,\rho_2}^{(a,b;c)} & = &  \mathcal{F}_{\rho_1,\rho_2}^{(a,b-1;c-1)} -C^{(a,b,c)}_{\rho_1}  \mathcal{F}_{\rho_1+1,\rho_2}^{(a,b-1;c-1)} -  \label{AppI4} \\
&&   C^{(a,b,c)}_{\rho_2}  \mathcal{F}_{\rho_1,\rho_2+1}^{(a,b-1;c-1)} 
  +C^{(a,b,c)}_{\rho_1} C^{(a,b,c)}_{\rho_2}  \mathcal{F}_{\rho_1+1,\rho_2+1}^{(a,b-1;c-1)} \,,  \nn \\
   \mathcal{F}_{\rho_1,\rho_2}^{(a,b;c)} & = &  \mathcal{F}_{\rho_1,\rho_2}^{(a-1,b;c-1)} -C^{(b,a,c)}_{\rho_1}  \mathcal{F}_{\rho_1+1,\rho_2}^{(a-1,b;c-1)}  -  \label{AppI5}\\
   && C^{(b,a,c)}_{\rho_2}  \mathcal{F}_{\rho_1,\rho_2+1}^{(a-1,b;c-1)} 
  +C^{(b,a,c)}_{\rho_1} C^{(b,a,c)}_{\rho_2}  \mathcal{F}_{\rho_1+1,\rho_2+1}^{(a-1,b;c-1)} \,,  \nn \\
\frac{1}{z\bar z}   \mathcal{F}_{\rho_1,\rho_2}^{(a,b;c)} & = &  \mathcal{F}_{\rho_1-1,\rho_2-1}^{(a+1,b+1;c+2)}  \,,  \label{AppI6} \\
(z-\bar z) L(a) \mathcal{F}_{\rho_1,\rho_2}^{(a,b;c)} & = &(\rho_2-\rho_1) \mathcal{F}_{\rho_1,\rho_2}^{(a,b-1;c-1)} -(\rho_1+\rho_2+c-1) C_{\rho_1}^{(a,b,c)}\mathcal{F}_{\rho_1+1,\rho_2}^{(a,b-1;c-1)}
+  \label{AppI7} \\
&& (\rho_1+\rho_2+c-1) C_{\rho_2}^{(a,b,c)}\mathcal{F}_{\rho_1,\rho_2+1}^{(a,b-1;c-1)} - (\rho_2-\rho_1) C_{\rho_1}^{(a,b,c)}C_{\rho_2}^{(a,b,c)}\mathcal{F}_{\rho_1+1,\rho_2+1}^{(a,b-1;c-1)}\,, \nn  \\
\frac{z-\bar z}{z\bar z}L(b) \mathcal{F}_{\rho_1,\rho_2}^{(a,b;c)} & = &(\rho_2-\rho_1) \mathcal{F}_{\rho_1-1,\rho_2-1}^{(a,b+1;c+1)} -(\rho_1+\rho_2+c-1) C_{\rho_1}^{(b,a,c)}\mathcal{F}_{\rho_1,\rho_2-1}^{(a,b+1;c+1)}
+  \label{AppI8} \\
&& (\rho_1+\rho_2+c-1) C_{\rho_2}^{(b,a,c)}\mathcal{F}_{\rho_1-1,\rho_2}^{(a,b+1;c+1)} - (\rho_2-\rho_1) C_{\rho_1}^{(b,a,c)}C_{\rho_2}^{(b,a,c)}\mathcal{F}_{\rho_1,\rho_2}^{(a,b+1;c+1)}\,.   \nn
\eea
The relations (\ref{AppI1})-(\ref{AppI3})  were first derived in ref.\cite{Dolan:2003hv} (see also ref.\cite{Dolan:2011dv}), while the relations (\ref{AppI7}) and (\ref{AppI8}) are novel to this paper.
It is straightforward to see that eqs.(\ref{eq:properties_group1})-(\ref{eq:properties_group3}) can be derived using proper combinations of eqs.(\ref{AppI1})-(\ref{AppI8}).
For instance, the action of the first term appearing in the r.h.s. of eq.(\ref{eq:casimir_modifed_part-}) is reproduced (modulo a trivial constant factor) by taking the combined action given by  ( (\ref{AppI2})$-$(\ref{AppI1}) )$ \times$  (\ref{AppI7})$\times$ (\ref{AppI4}).
All other terms in eqs.(\ref{eq:casimir_modifed_part0})-(\ref{eq:casimir_modifed_part-}) are similarly deconstructed.

\section{The Conformal Blocks for $p=1$}
\label{app:peq1}

We report in this appendix the full explicit solution for the two conformal blocks $G_0^{(1)}$ and $G_1^{(1)}$ associated to the exchange of fermion operators
of the kind ${\cal O}^{(\ell,\ell+1)}$ for the specific values
\be
a = \frac 12\,, \ \ \ \ b =-\frac 12 \,.
\ee
We choose as undetermined coefficient $c_{0,-1}^1$ and report below the values of the coefficients normalized to $c_{0,-1}^1$. 
We have
\be\label{cp1Some}
c^0_{-2,0}= \frac{(2+\ell) }{2\, (1+\ell)},\;\;
c^0_{-1,-1}= -\frac{\ell\, }{2\, (1+\ell)},\;\;
c^1_{-1,0}= -\frac{(3+\ell)}{1+\ell}.
\ee
\bea
\nn c^0_{-1,0} &=& \frac{(3+\ell) (-1+2 \Delta ) (-1+2 \ell+2 \Delta ) }{8 (1+\ell) (-3+2 \Delta ) (1+2 \ell+2 \Delta )},\\
\nn c^0_{-1,1} &=& -\frac{(2+\ell) (5+2 \ell-2 \Delta )^2 (-7+2 \Delta ) }{32 (1+\ell) (3+2 \ell-2 \Delta ) (7+2 \ell-2 \Delta ) (-3+2 \Delta )},\\
\nn c^0_{0,-1}  &=& -\frac{(-1+2 \Delta ) (-1+2 \ell+2 \Delta )}{8 (-3+2 \Delta ) (1+2 \ell+2 \Delta )},\\
\nn c^0_{0,0}  &=& \frac{\ell (-7+2 \Delta ) (-1+2 \ell+2 \Delta )^2 }{32 (1+\ell) (-3+2 \Delta ) (-3+2 \ell+2 \Delta ) (1+2 \ell+2 \Delta )},\\
\nn c^0_{0,1}  &=& -\frac{(3+\ell) (5+2 \ell-2 \Delta )^2 (-5+2 \Delta ) (-1+2 \ell+2 \Delta ) }{128 (1+\ell) (3+2 \ell-2 \Delta ) (7+2 \ell-2 \Delta) (-3+2 \Delta ) (1+2 \ell+2 \Delta )},\\
\nn c^0_{1,0}  &=& \frac{(-5+2 \Delta ) (-1+2 \ell+2 \Delta ) (3+2 \ell+2 \Delta )^2}{128 (-3+2 \Delta ) (1+2 \ell+2 \Delta )^2 (5+2 \ell+2\Delta )},\\
\nn c^1_{-1,1} &=& -\frac{(2+\ell) (5+2 \ell-2 \Delta ) (-1+2 \Delta )}{4 (1+\ell) (7+2 \ell-2 \Delta ) (-3+2 \Delta )},\\
\nn c^1_{0,2}   &=& \frac{(2+\ell) (1+2 \ell-2 \Delta ) (5+2 \ell-2 \Delta )^2 (-5+2 \Delta )}{64 (1+\ell) (3+2 \ell-2 \Delta )^2 (7+2 \ell-2 \Delta) (-3+2 \Delta )},\\
\nn c^1_{1,0}  &=&  -\frac{(-7+2 \Delta ) (-1+2 \ell+2 \Delta ) (3+2 \ell+2 \Delta ) }{16 (-3+2 \Delta ) (1+2 \ell+2 \Delta )^2},\\
\nn c^1_{1,1}  &=& -\frac{\ell (5+2 \ell-2 \Delta ) (-5+2 \Delta ) (-1+2 \ell+2 \Delta ) (3+2 \ell+2 \Delta )}{64 (1+\ell) (7+2 \ell-2 \Delta ) (-3+2\Delta ) (1+2 \ell+2 \Delta )^2}
\eea
\bea
c^1_{0,0} & = & \frac{1}{4 (1+\ell) (11+2 \ell-2 \Delta ) (-3+2 \Delta ) (-3+2 \ell+2 \Delta ) (1+2 \ell+2 \Delta)} \times  \nn \\
&&  \bigg(576-384 \Delta + \ell \Big(627 -2\ell  (-29+2 \ell (7+2 \ell))-472 \Delta +4 \ell (-47+4 \ell (3+\ell)) \Delta \nn \\
&&+8 (-9+\ell (19+2 \ell)) \Delta ^2-16 (-6+\ell) \Delta ^3-16 \Delta ^4\Big)\bigg)\,, \nn \\
 c^1_{0,1}& = & \frac{(5+2 \ell-2 \Delta ) }{16 (1+\ell) (3+2 \ell-2 \Delta ) (7+2 \ell-2 \Delta ) (-3+2 \Delta
) (-3+2 \ell+2 \Delta ) (1+2 \ell+2 \Delta )}\times \nn \\
&& \bigg(\ell (643-14 \ell (-3+2 \ell (9+2 \ell)))+4 \ell (-232+\ell (-115+4 \ell (1+\ell))) \Delta +8 (3+\ell)  \nn \\
&& (-24+\ell (17+2 \ell)) \Delta ^2-16 (-7+\ell) (3+\ell) \Delta ^3-16 (3+\ell) \Delta ^4+27 (9+4 \Delta )\bigg)\,. \nn 
\eea
The asymptotic behaviour of the CBs for $z,\bar z\rightarrow 0$ ($z\rightarrow 0$ first) is dominated by the coefficients with $n=-1$ and the lowest value of $m$, i.e. $c_{-1,-1}^0$ and $c_{0,-1}^1$.
For $\ell=0$, the asymptotic behaviour of $G_0^{(1)}$ is given by the next term $c_{0,-1}^0$, since $c_{-1,-1}^0$ in eq.(\ref{cp1Some}) vanishes. This in agreement with the asymptotic behaviour of the CBs found in subsection \ref{subsec:limitz0}.
Notice how the complexity of the $c_{m,n}^e$ varies from coefficient to coefficient. In general the most complicated ones are those in the ``interior" of the octagons (hexagons only for $p=1$).

\end{document}